\documentclass{article}

\usepackage[utf8]{inputenc}
\usepackage{amsmath,amsfonts,amsthm} 
\usepackage{mathtools}
\usepackage[a4paper, left=1.45cm, right=1.4cm, top=1.2cm, bottom=1.7cm,]{geometry}
\usepackage[backend=biber, style=apa]{biblatex}
\usepackage{float}
\usepackage{tikz}

\usepackage{caption}
\usepackage{subcaption}
\usepackage{authblk}
\usepackage{graphicx}
\usepackage[title]{appendix}

\providecommand{\keywords}[1]
{
  \small  
  \textbf{\textit{Keywords---}} #1
}

\bibliography{bibliography}

\title{Bayesian D- and I-optimal designs for choice experiments involving mixtures and process variables}

\author{Mario Becerra$^1$}
\author{Peter Goos$^{1,2}$}
\affil{$^1$ Faculty of Bioscience Engineering, KU Leuven, Leuven, Belgium}
\affil{$^2$ Faculty of Business and Economics, Universiteit Antwerpen, Antwerpen, Belgium}

\date{\vspace{-5ex}} 

\newcommand{\set}[1]{\left\{ #1 \right\}}

\begin{document}

\maketitle

\begin{abstract}
Many food products involve mixtures of ingredients, where the mixtures can be expressed as combinations of ingredient proportions. In many cases, the quality and the consumer preference may also depend on the way in which the mixtures are processed. The processing is generally defined by the settings of one or more process variables. Experimental designs studying the joint impact of the mixture ingredient proportions and the settings of the process variables are called mixture-process variable experiments. In this article, we show how to combine mixture-process variable experiments and discrete choice experiments, to quantify and model consumer preferences for food products that can be viewed as processed mixtures. First, we describe the modeling of data from such combined experiments. Next, we describe how to generate D- and I-optimal designs for choice experiments involving mixtures and process variables, and we compare the two kinds of designs using two examples.
\end{abstract}
\keywords{Choice experiment; I-optimality; Mixture coordinate-exchange algorithm; Mixture-process variable experiment; Multinomial logit model; Scheffé models}


\section{Introduction}

As pointed out in the review paper on the state of the art of discrete choice experiments in food research by \textcite{Lizin2022state}, there has been a steady increase in the number of publications on the use of discrete choice experiments concerning food since 2000. A large number of discrete choice experiments in these papers deal with food safety or safety risks, origin or traceability, health or nutrition, biotechnology or genetic modification and animal welfare. The product categories mainly involved meat (beef, pork, poultry, and processed meat products), organic foods, functional foods and foods with nutrition or health claims. In recent years, alternatives to conventional meat received increasing attention. \textcite{Lizin2022state} also mention that a limited number of choice experiments in published papers were concerned with wine, olive oil, eggs and vegetables.

Despite the fact that many food products involve mixtures of ingredients, publications concerning food-related choice experiments with mixtures are scarce. The first known application of a discrete choice experiment concerning mixtures was published by \textcite{courcoux1997methode}, who modeled the preferences for cocktails involving different proportions of mango juice, lime juice, and blackcurrant syrup. \textcite{goos_hamidouche_2019_choice} defined a way to combine Scheffé models for data from mixture experiments with the logit type models typically used for choice experiments, and presented an alternative analysis of the data from \textcite{courcoux1997methode}. \textcite{ruseckaite_bayesian_2017} and \textcite{becerra2021bayesian} demonstrated how D- and I-optimal designs can be generated for choice experiments with mixtures, applied their work to the cocktail experiment and used an additional example concerning a sports drink. 

As witnessed by many of the examples in \textcite{cornell2002experiments}, the quality of food products involving mixtures of ingredients often also depends on characteristics unrelated to the composition of the mixture. For example, the firmness of a fish patty depends not only on the types of fish used, but also on baking temperature, baking time, and frying time.
The color, aroma, taste, texture and mouthfeel of \textit{pastillas de leche}, a popular Filipino candy, depend on baking time and temperature in addition to mixture ingredients such as cornstarch, flour, glucose, sugar and milk \autocite{apelladooptimization}.
The aroma, hardness, crispness, color and fracture force of apple biscuits are affected by the mixture ingredients and the microwave blanching of the apples \autocite{skaltsi2022development}. In the general literature on mixture experiments, variables such as baking temperature, baking time, frying time, serving temperature, and microwave blanching are typically called \textit{process variables} \autocite{goos_jones_optimal_2011}.

The fact that the quality of food products involving mixtures depends on the settings of such process variables implies that consumer preferences for these kinds of products will also be impacted by the process variables' settings. For this reason, in this article, we develop the methodology required to perform discrete choice experiments involving mixtures as well as process variables. First, we present a parsimonious model for data from choice experiments with mixture and process variables. Next, we discuss how to generate D- and I-optimal designs for such choice experiments. We discuss D-optimal designs because the D-optimality criterion is the most popular criterion for designing choice experiments; and I-optimal designs because they focus on precise predictions and precise predictions are helpful to find the optimal mixture formulation in combination with optimal settings for the process variables.

The rest of the paper is organized as follows. In Section \ref{sec:models}, we introduce the most often used models for mixture experiments with process variables, the multinomial logit model for choice data and the combination of these two to model choice data concerning mixtures and process variables. In Section \ref{sec:opt_des_criteria}, we discuss the two most commonly used metrics to measure the quality of experimental designs. In Section \ref{sec:results}, we present some of our computational results and provide example designs for two choice experiments involving a mixture and one or more process variables. Finally, in Section \ref{sec:discussion}, we summarize our work and sketch possible directions for future research.


\section{Models}\label{sec:models}

In this section, we introduce the most commonly used models for data from mixture experiments with process variables as well as the multinomial logit model for choice data, and explain how to combine the two models for data from choice experiments involving mixtures and process variables.

\subsection{Models for data from mixture experiments including process variables}
\label{sectionwithconstraints}

Mixture experiments involve two or more ingredients and a response variable that depends only on the relative proportions of the ingredients in the mixture. Each mixture is described as a combination of $q$ ingredient proportions, with the constraint that these proportions sum up to one. Due to this constraint, a classical regression model involving an intercept and linear terms in the ingredient proportions exhibits perfect collinearity. Therefore, researchers must use dedicated regression models when analyzing data from mixture experiments. The most commonly used family of models for data from mixture experiments is the Scheffé family (\cite{scheffe1958experiments}; \cite{scheffe1963simplex}). The most popular Scheffé models are the first-order, second-order, and special-cubic models. 

Denoting the response in a traditional mixture experiment with a continuous outcome by $Y$ and the $q$ ingredient proportions by $x_1,x_2,\dots,x_q$, with $x_i \geq 0$ and $\sum_{i = 1}^q x_i = 1$, the first-order Scheffé model is
\begin{equation}\label{eq_scheffe_first_order}
    Y = \sum_{i = 1}^q \beta_i x_i + \varepsilon.
\end{equation}
The second-order Scheffé model is
\begin{equation}\label{eq_scheffe_second_order}
    Y = 
      \sum_{i = 1}^q \beta_i x_i + 
      \sum_{i = 1}^{q-1} \sum_{j = i+1}^q \beta_{ij} x_i x_j + 
      \varepsilon,
\end{equation}
and, finally, the special-cubic Scheffé model is
\begin{equation}\label{eq_scheffe_special_cubic}
    Y = 
      \sum_{i = 1}^q \beta_i x_i + 
      \sum_{i = 1}^{q-1} \sum_{j = i+1}^q \beta_{ij} x_i x_j + 
      \sum_{i = 1}^{q-2} \sum_{j = i+1}^{q-1} \sum_{k = j+1}^{q} \beta_{ijk} x_i x_j x_k + 
      \varepsilon.
\end{equation}
In all three cases, $\varepsilon$ denotes the error term, which, for continuous outcomes, is typically assumed to be normally distributed.

In certain experiments involving mixtures, additional factors that might affect the response are studied as well. Generally, these factors describe how the mixture is processed (where the word `processed' should be interpreted in a broad sense). These additional factors are therefore referred to as process variables, and the resulting experiments are called mixture-process variable experiments. For instance, a dough needs to be baked at a certain temperature for a certain time, while the cocktails from the example in Section~\ref{cex} need to be cooled to a certain temperature before being served, and the fish patties from the example in Section~\ref{fex} are cooked and fried for a specific time at a specific temperature.

Models that involve $q$ mixture ingredients and $r$ process variables can be obtained by combining Scheffé models for the ingredient proportions with response surface models for the process variables (\cite{cornell1984fractional}; \cite{cornell1988analyzing}; \cite{kowalski2000new}; \cite{goos_jones_optimal_2011}). For example, consider the second-order Scheffé model in Equation~(\ref{eq_scheffe_second_order}) for $q$ ingredients $x_1, x_2, \dots, x_q$ and a main-effects-plus-two-factor-interaction model for $r$ process variables $z_1, z_2, \dots, z_r$ defined as
\begin{equation}\label{eq_response_surf_model}
    Y = 
      \alpha_0 +
      \sum_{k = 1}^r \alpha_k z_k + 
      \sum_{k = 1}^{r-1} \sum_{l = k+1}^r \alpha_{kl} z_k z_l + 
      \varepsilon.
\end{equation}
One combined model crosses the terms in Equation~(\ref{eq_scheffe_second_order}) with each of those in Equation~(\ref{eq_response_surf_model}):
\begin{equation}\label{eq_full_combined_model}
    \begin{aligned}
    Y = \sum_{i = 1}^q \beta_i x_i
      + \sum_{i = 1}^{q-1} \sum_{j = i+1}^q \beta_{ij} x_i x_j
      + \sum_{i = 1}^{q} \sum_{k = 1}^r \gamma_{ik} x_i z_k
      + \sum_{i = 1}^q \sum_{k = 1}^{r-1} \sum_{l = k+1}^r \gamma_{ikl} x_i z_k z_l \\
      + \sum_{i = 1}^{q-1} \sum_{j = i+1}^q \sum_{k = 1}^r \delta_{ijk} x_i x_j z_k
      + \sum_{i = 1}^{q-1} \sum_{j = i+1}^q \sum_{k = 1}^{r-1} \sum_{l = k+1}^r \delta_{ijkl} x_i x_j z_k z_l
      + \varepsilon.
    \end{aligned}
\end{equation}
This model allows the effects of both the ingredient proportions and process variables to jointly affect the response variable. In other words, the model allows the effects of the process variables to depend on the ingredient proportions and the effects of the ingredient proportions to depend on the process variables. The combined model in Equation~(\ref{eq_full_combined_model}) does not include any main effects of the process variables $z_1, \dots, z_r$. This is because their inclusion would result in an inestimable model due to perfect collinearity. In the event that the effects of the process variables do not depend on the ingredient proportions, all $\gamma_{ik}$ as well as all $\gamma_{ikl}$ in the combined model are equal and all 
$\delta_{ijk}$ and all $\delta_{ijkl}$ are zero. In such event, the model simplifies to
\begin{equation}\label{eq_combined_model_2}
    Y = \sum_{i = 1}^q \beta_i x_i
      + \sum_{i = 1}^{q-1} \sum_{j = i+1}^q \beta_{ij} x_i x_j
      + \sum_{k = 1}^r \alpha_k z_k
      + \sum_{k = 1}^{r-1} \sum_{l = k+1}^r \alpha_{kl} z_k z_l
      + \varepsilon.
\end{equation}
This alternative model also combines the models in Equations (\ref{eq_scheffe_second_order}) and (\ref{eq_response_surf_model}), but without crossing any of the terms. Depending on the application, it may be necessary to extend the above models by including cubic terms involving the mixture ingredient proportions (as in the special-cubic Scheffé model in Equation~\eqref{eq_scheffe_special_cubic}) or quadratic terms in the process variables. An example of such an extended model would be
\begin{equation}\label{eq_full_combined_modelnew}
    \begin{aligned}
    Y = \sum_{i = 1}^q \beta_i x_i
      + \sum_{i = 1}^{q-1} \sum_{j = i+1}^q \beta_{ij} x_i x_j + \sum_{i = 1}^{q-2} \sum_{j = i+1}^{q-1} \sum_{k = j+1}^{q} \beta_{ijk} x_i x_j x_k
      + \sum_{i = 1}^{q} \sum_{k = 1}^r \gamma_{ik} x_i z_k
      + \sum_{i = 1}^q \sum_{k = 1}^{r-1} \sum_{l = k+1}^r \gamma_{ikl} x_i z_k z_l \\
      + \sum_{i = 1}^{q-1} \sum_{j = i+1}^q \sum_{k = 1}^r \delta_{ijk} x_i x_j z_k
      + \sum_{i = 1}^{q-1} \sum_{j = i+1}^q \sum_{k = 1}^{r-1} \sum_{l = k+1}^r \delta_{ijkl} x_i x_j z_k z_l + \sum_{i = 1}^r \alpha_i z_i^2
      + \varepsilon.
    \end{aligned}
\end{equation}

A problem with the combined model in Equation~(\ref{eq_full_combined_model}) is that its number of parameters quickly increases with the number of mixture ingredients and process variables: for $q$ mixture ingredients and $r$ process variables, the total number of parameters is $[q + q(q - 1)/2] \times [1 + r + r(r - 1)/2]$. The extended model in Equation~(\ref{eq_full_combined_modelnew}) even involves $q(q-1)(q-2)/6+r$ extra parameters. In contrast, the model described in Equation~(\ref{eq_combined_model_2}) involves a number of parameters that is as low as $[q + q(q - 1)/2]+[r + r(r - 1)/2]$. The drawback of the latter model is that it may not be realistic. For this reason, \textcite{kowalski2000new} suggest a compromise model involving $q + q(q-1)/2 + qr + r(r-1)/2 + r$ terms:
\begin{equation}
    \label{eq:scheffe_second_order_pv}
    Y = 
      \sum_{k = 1}^q \gamma_k^0 x_k + 
      \sum_{k = 1}^{q-1} \sum_{l = k+1}^q \gamma_{kl}^0 x_k x_l + 
      \sum_{i = 1}^{r} \sum_{k = 1}^q \gamma_{k}^i x_k z_i + 
      \sum_{i = 1}^{r-1} \sum_{j = i + 1}^r \alpha_{ij} z_i z_j + 
      \sum_{i = 1}^r \alpha_i z_i^2 + 
      \varepsilon.
\end{equation}
Because this compromise model strikes a balance between the overly complex models in Equations~(\ref{eq_full_combined_model}) and~(\ref{eq_full_combined_modelnew}) and the overly simple model in Equation~(\ref{eq_combined_model_2}), we use it as our starting point for computing optimal designs for choice experiments involving mixtures and process variables in the remainder of this paper.

\subsection{Multinomial logit model for choice data}

The multinomial logit model builds on random-utility theory and assumes that a respondent in a choice experiment faces $S$ choice sets involving $J$ alternatives. The model assumes that, within each choice set $s \in \set{1, ..., S}$, each respondent chooses the alternative that has the highest perceived utility. Therefore, the probability that a respondent chooses alternative $j \in \set{1, ..., J}$ in choice set $s$, denoted by $p_{js}$, is the probability that the perceived utility of alternative $j$ in choice set $s$, denoted by $U_{js}$, is larger than that of the other alternatives in the choice set:
\begin{equation*}
    \label{eq_p_js_max}
    p_{js} = \mathbb{P} \left[ U_{js} > \max(U_{1s}, ..., U_{j-1, s}, U_{j+1, s}, ..., U_{Js} ) \right].
\end{equation*}
Since, generally, each alternative in a choice set has a set of observable attributes that characterize it, the perceived utility $U_{js}$ can be expressed as 
\begin{equation}
    \label{eq_utility_js}
    U_{js} = \boldsymbol{f}^T(\boldsymbol{a}_{js}) \boldsymbol{\theta} + \varepsilon_{js},
\end{equation}
where $\boldsymbol{a}_{js}$ is the vector that contains the attributes corresponding to alternative $j$ in choice set $s$, $\boldsymbol{f}(\boldsymbol{a}_{js})$ represents the model expansion of this attribute vector, and $\boldsymbol{\theta}$ is the vector containing the model parameters. The model parameters contained within $\boldsymbol{\theta}$ express the preferences of the respondents for the alternatives' attributes. In the multinomial logit model, the error terms $\varepsilon_{js}$ are assumed to be independent and identically Gumbel distributed. The Gumbel distribution is also known as the generalized extreme value distribution of type I and as the log-Weibull distribution. As a result of the distributional assumption, it can be shown that
\begin{equation}
    \label{eq_p_j}
    p_{js} = \frac{ \exp{ \left[ \boldsymbol{f}^T(\boldsymbol{a}_{js}) \boldsymbol{\theta} \right]} }{ \sum_{t = 1}^J \exp{ \left[ \boldsymbol{f}^T(\boldsymbol{a}_{ts}) \boldsymbol{\theta} \right]} }.
\end{equation}

\subsection{Model for choice data concerning mixtures and process variables}\label{subsec:mixtures_and_mnl_model}

In this paper, we focus on choice experiments involving mixtures and process variables. Therefore, we assume that the attributes of the alternatives in the experiments are the proportions of the ingredients of a mixture and the settings of the process variables. Consequently, we assume that the attribute vector $\boldsymbol{a}_{js}$ from Equation~(\ref{eq_utility_js}) contains the $q$ ingredient proportions $x_1,x_2,\dots,x_q$ and the $r$ process variables $z_1,\dots,z_r$ of the $j$-th alternative in choice set $s$ and that $\boldsymbol{f}(\boldsymbol{a}_{js})$ represents the model expansion of these proportions and process variables. As a proof of concept, in this paper we base the polynomial expansion $\boldsymbol{f}(\boldsymbol{a}_{js})$ on a model combining a second-order Scheffé model for the $q$ ingredients in the mixture with a main-effects-plus-two-factor-interaction model for the $r$ process variables, as in Equation~(\ref{eq:scheffe_second_order_pv}).

When starting from the main-effects-plus-two-factor-interaction model in Equation~(\ref{eq:scheffe_second_order_pv}), the most natural thing to do would be to write the perceived utility $U_{js}$ of alternative $j$ in choice set $s$ as
\begin{equation*}
    U_{js} =
    \sum_{i = 1}^{q} \gamma_i^{0} x_{ijs} + 
    \sum_{i = 1}^{q-1} \sum_{k = i+1}^q \gamma_{ik}^0 x_{ijs} x_{kjs} + 
    \sum_{i = 1}^{r} \sum_{k = 1}^q \gamma_{k}^i x_{kjs} z_{ijs} + 
    \sum_{i = 1}^{r-1} \sum_{k = i + 1}^r \alpha_{ik} z_{ijs} z_{kjs} + 
    \sum_{i = 1}^r \alpha_i z_{ijs}^2 +
    \varepsilon_{js},
\end{equation*}
where $x_{ijs}$ denotes the proportion of the $i$-th mixture ingredient in alternative $j$ from choice set $s$, and $z_{kjs}$ denotes the setting of the $k$-th process variable for alternative $j$ in choice set $s$, and the error terms $\varepsilon_{js}$ are assumed to be independent and identically Gumbel distributed. However, as explained by \textcite{ruseckaite_bayesian_2017}, \textcite{goos_hamidouche_2019_choice}, and \textcite{becerra2021bayesian}, due to the constraint that the ingredient proportions sum up to one, this leads to an inestimable multinomial logit model. As a consequence of the constraint, we can rewrite $x_{qjs}$ as $1 - x_{1js} - ... - x_{q-1,js}$ and $U_{js}$ as
\begin{equation*}
    \label{eq_u_js_mixt_unidetif_pv}
    \begin{aligned}
    U_{js} 
    &= 
    \sum_{i = 1}^{q-1} \gamma_i^{0} x_{ijs} +
    \gamma_q^{0} (1 - x_{1js} - ... - x_{q-1,j,s}) + 
    \sum_{i = 1}^{q-1} \sum_{k = i+1}^q \gamma_{ik}^0 x_{ijs} x_{kjs} + 
    \sum_{i = 1}^{r} \sum_{k = 1}^q \gamma_{k}^i x_{kjs} z_{ijs} + 
    \sum_{i = 1}^{r-1} \sum_{k = i + 1}^r \alpha_{ik} z_{ijs} z_{kjs} + 
    \sum_{i = 1}^r \alpha_i z_{ijs}^2 +
    \varepsilon_{js} \\ 
    &=
    \gamma_q^{0} + 
    \sum_{i = 1}^{q-1} (\gamma_i^{0} - \gamma_q^{0}) x_{ijs} +
    \sum_{i = 1}^{q-1} \sum_{k = i+1}^q \gamma_{ik}^0 x_{ijs} x_{kjs} + 
    \sum_{i = 1}^{r} \sum_{k = 1}^q \gamma_{k}^i x_{kjs} z_{ijs} + 
    \sum_{i = 1}^{r-1} \sum_{k = i + 1}^r \alpha_{ik} z_{ijs} z_{kjs} + 
    \sum_{i = 1}^r \alpha_i z_{ijs}^2 +
    \varepsilon_{js}.
    \end{aligned}
\end{equation*}
This final expression for the perceived utility $U_{js}$ involves a constant, $\gamma_q^{0}$. Since the multinomial logit model only takes into account differences in utility, that constant causes the model to be ill-defined and, hence, inestimable. This can be circumvented by dropping $\gamma_q^{0}$, defining the parameters
$\gamma_i^{0*} = \gamma_i^{0} - \gamma_q^{0}$ for $i \in \set{1, ..., q-1}$, and using the following expression for the perceived utility:
\begin{equation}
    \label{eq:utility_js_pv}
    U_{js} =
    \sum_{i = 1}^{q-1} \gamma_i^{0*} x_{ijs} + 
    \sum_{i = 1}^{q-1} \sum_{k = i+1}^q \gamma_{ik}^0 x_{ijs} x_{kjs} + 
    \sum_{i = 1}^{r} \sum_{k = 1}^q \gamma_{k}^i x_{kjs} z_{ijs} + 
    \sum_{i = 1}^{r-1} \sum_{k = i + 1}^r \alpha_{ik} z_{ijs} z_{kjs} + 
    \sum_{i = 1}^r \alpha_i z_{ijs}^2 +
    \varepsilon_{js}.
\end{equation}
The parameter vector $\boldsymbol{\theta}$ then becomes
$$
    \boldsymbol{\theta} = 
    \left(
    \gamma_1^{0*}, \gamma_2^{0*}, ..., \gamma_{q-1}^{0*}, \gamma_{1,2}^{0}, ..., \gamma_{q-1,q}^{0}, \gamma_{1}^1, \gamma_{2}^1, ..., \gamma_{q}^1, \gamma_{q}^2, ..., \gamma_{q}^r, \alpha_{1,2}, ..., \alpha_{r-1,r}, \alpha_{1}, ..., \alpha_{r}
    \right)^T.
$$
This vector has $q + \frac{q(q-1)}{2} + qr + \frac{r(r-1)}{2} + r - 1$ elements. 


\section{Optimal design criteria}
\label{sec:opt_des_criteria}

In the literature on the optimal design of choice experiments in general, several criteria have been studied. \textcite{kessels2006comparison} elaborate on the D-, I-, A-, and G-optimality criteria for the multinomial logit model and compare the performances of the resulting choice designs. However, in the literature on optimal design of choice experiments with mixtures, the two optimality metrics that have been studied are D-optimality and I-optimality. In this section, we extend the D- and I-optimality criteria to cope with the multinomial logit model for choice experiments involving mixtures as well as process variables.

\subsection{Information matrix}

In order to create D- and I-optimal experimental designs, we need to compute a design's information matrix corresponding to the model under investigation. For the multinomial logit model, the information matrix depends on the unknown parameter vector $\boldsymbol{\theta}$ through the choice probabilities $p_{js}$ defined in Equation~{(\ref{eq_p_j})}. 
This is typical for models that are not linear in the parameters, such as discrete choice models, and it implies that prior information is needed to find optimal designs. This information can be provided in the form of a point estimate, or in the form of a prior distribution (\cite{atkinson199614}; \cite{kessels2006comparison}; \cite{ruseckaite_bayesian_2017}; \cite{becerra2021bayesian}). The use of a point estimate leads to so-called locally optimal designs, which have the problem that they may perform poorly for values of the parameter vector $\boldsymbol{\theta}$ for which they were not optimized. This weakness of locally optimal designs is, of course, highly relevant given that the true values of the model parameters are unknown. An alternative is to use a prior distribution, which leads to so-called Bayesian optimal designs. In addition to taking into account prior information, Bayesian optimal designs also take into account the uncertainty about the parameter vector $\boldsymbol{\theta}$ through the use of a prior distribution $\pi(\boldsymbol{\theta})$ that summarizes the prior knowledge concerning the parameter vector $\boldsymbol{\theta}$.

The information matrix $\boldsymbol{I}(\boldsymbol{X}, \boldsymbol{\theta})$ for the multinomial logit model is the sum of the information matrices of each of the $S$ choice sets \autocite{kessels2006comparison}:
\begin{equation*}
    \label{eq_inf_matrix}
    \boldsymbol{I}(\boldsymbol{X}, \boldsymbol{\theta}) = 
        \sum_{s = 1}^S 
        \boldsymbol{X}_s^T (\boldsymbol{P}_s - \boldsymbol{p}_s \boldsymbol{p}_s^T) \boldsymbol{X}_s,
\end{equation*}
with $\boldsymbol{p}_s = \left( p_{1s}, ..., p_{Js} \right)^T$, $\boldsymbol{P}_s = \mathrm{diag}(\boldsymbol{p}_s)$, $\boldsymbol{X}_s^T = \left[ \boldsymbol{f}(\boldsymbol{a}_{1s}), \boldsymbol{f}(\boldsymbol{a}_{2s}), ..., \boldsymbol{f}(\boldsymbol{a}_{Js}) \right]$  the model matrix containing the model expansions of the attribute levels of all $J$ alternatives in choice set $s$, and $\boldsymbol{X} = \left[ \boldsymbol{X}_1, ..., \boldsymbol{X}_S \right]$ the model matrix for all $S$ choice sets. The inverse of the information matrix is the asymptotic variance-covariance matrix of the maximum likelihood estimates of the  parameter vector $\boldsymbol{\theta}$.

\subsection{D-optimal designs}

For a model matrix $\boldsymbol{X}$ and prior parameter vector $\boldsymbol{\theta}$, the D-optimality criterion can be defined as
\begin{equation}\label{eq_d_optim_classical}
    \mathcal{D} =
    \left[ \det \left( 
            \boldsymbol{I}^{-1}(\boldsymbol{X}, \boldsymbol{\theta})
         \right) \right]^{\frac{1}{m}},
\end{equation}
where $\boldsymbol{I}^{-1}(\boldsymbol{X}, \boldsymbol{\theta})$ is the inverse of the information matrix and $m$ is the number of parameters in the model. A D-optimal design minimizes the $\mathcal{D}$-value. Since the D-optimal design approach focuses on minimizing the generalized variance of the maximum likelihood estimators of the model parameters, it can be viewed as an estimation-based approach. D-optimality is arguably the most traditional metric used in the literature on the design of choice experiments (\cite{bliemer2009efficient}; \cite{bliemer2010construction}; \cite{bliemer2011experimental}; \cite{burgess2005optimal}; \cite{grasshoff2003optimal}; \cite{kessels_usefulness_2011}).

The definition in Equation~(\ref{eq_d_optim_classical}) uses a prior point estimate of the parameter vector $\boldsymbol{\theta}$. However, as we mentioned above, a prior distribution can also be used to obtain a Bayesian D-optimal design. The Bayesian D-optimality criterion is generally defined in the literature as the average of the D-optimality criterion over the prior distribution (\cite{bliemer2009efficient}; \cite{bliemer2011experimental}; \cite{kessels_usefulness_2011}). Therefore, following \textcite{ruseckaite_bayesian_2017} and \textcite{becerra2021bayesian}, we define the Bayesian D-optimality criterion for the multinomial logit model as
\begin{equation}
    \label{eq_pseudo_bayes_d_eff}
    \mathcal{D}_B =
        \int_{\mathbb{R}^{m}}
        \left[  
            \det \left( 
                \boldsymbol{I}^{-1}(\boldsymbol{X}, \boldsymbol{\theta})
             \right)
            \right]^{\frac{1}{m}}
        \pi(\boldsymbol{\theta}) d\boldsymbol{\theta},
\end{equation}
\noindent
where $\pi(\boldsymbol{\theta})$ is the prior distribution of $\boldsymbol{\theta}$. Note that we call a design that minimizes the expression in Equation~(\ref{eq_pseudo_bayes_d_eff}) a Bayesian D-optimal design, even though the criterion does not take into account the posterior distribution and some authors therefore prefer to call such a design a pseudo-Bayesian design (e.g., \textcite{ryan2016review}).

\subsection{I-optimal designs}

The I-optimality criterion is generally defined as the average prediction variance over the experimental region, which is why it can be seen as a prediction-oriented criterion: it focuses on getting precise predictions with the estimated statistical model. I-optimality is also sometimes called V-optimality (\cite{gosy}; \cite{kessels2006comparison}).

When using choice models, there are two ways in which we can define I-optimality. If the goal is to predict choice probabilities, the I-optimality criterion is the average variance of the predicted choice probabilities. If the goal is to predict perceived utilities, the I-optimality criterion is the average variance of the predicted utilities. \textcite{becerra2021bayesian} introduced a computationally efficient definition for I-optimal designs for choice experiments focused on the perceived utilities. This is the definition we will use here too. Under this definition, the I-optimality criterion is
\begin{equation}\label{eq_i_eff}
    \mathcal{I} =
    \mathrm{tr}\left[ \boldsymbol{I}^{-1}(\boldsymbol{X}, \boldsymbol{\theta}) \boldsymbol{W}\right],
\end{equation}
where $\boldsymbol{I}^{-1}(\boldsymbol{X}, \boldsymbol{\theta})$ again denotes the inverse of the information matrix for model matrix $\boldsymbol{X}$ and prior parameter vector $\boldsymbol{\theta}$. The matrix $\boldsymbol{W}$ is the moments matrix, defined as
\begin{equation}    \label{eq_moments_matrix_utility}
    \boldsymbol{W} = \int_{\chi}  \boldsymbol{f}(\boldsymbol{a}_{js}) \boldsymbol{f}^T(\boldsymbol{a}_{js}) d\boldsymbol{a}_{js},
\end{equation}
with $\boldsymbol{f}(\boldsymbol{a}_{js})$ again the model expansion of attribute vector $\boldsymbol{a}_{js}$ and $\chi$ the experimental region which combines the $(q-1)$-dimensional simplex $S_{q-1}$ for the ingredient proportions and an $r$-dimensional hyperrectangle for the possible settings of the process variables. 

To compute the moments matrix $\boldsymbol{W}$ for the model described in Equation~(\ref{eq:utility_js_pv}), we first need to compute the matrix $\boldsymbol{f}(\boldsymbol{a}_{js}) \boldsymbol{f}^T(\boldsymbol{a}_{js})$, which has elements of the form
\begin{equation*}
    \left( \prod_{k = 1}^q x_k^{n_k} \right) \left( \prod_{l = 1}^r z_l^{m_l} \right),
\end{equation*}
for some $n_k, m_l \in \mathbb{N}$, $k \in \left\{ 1, ..., q \right\}$ and $l \in \left\{ 1, ..., r \right\}$.
Hence, each element of the moments matrix is of the form
\begin{equation*}
    \int_{\chi} \left( \prod_{k = 1}^q x_k^{n_k} \right) \left( \prod_{l = 1}^r z_l^{m_l} \right) dx_1 \dots dx_{q} dz_1 \dots dz_r,
\end{equation*}
which can be separated in two parts: one corresponding to the process variables and one part corresponding to the ingredient proportions. Therefore, assuming that the $r$ process variables take values from the intervals $\left[ a_1, b_1 \right]$, $\left[ a_2, b_2 \right]$, \dots, $\left[ a_r, b_r \right]$, the $i$-th element in the $j$-th column of the moments matrix, denoted by $W_{ij}$, can be calculated as
\begin{equation*}
    \label{moments_matrix1}
    \begin{aligned}
    W_{ij} 
    &= 
    \int_{a_1}^{b_1} \int_{a_2}^{b_2} ... \int_{a_r}^{b_r} \prod_{l = 1}^r z_l^{m_l} \left( \int_{S_{q-1}} \prod_{k = 1}^q x_k^{n_k} dx_1 ... dx_{q-1} \right) dz_1 \dots dz_r, \\ 
    &=
    \int_{a_1}^{b_1} \int_{a_2}^{b_2} ... \int_{a_r}^{b_r} \prod_{l = 1}^r z_l^{m_l} \left( \frac{\prod_{k = 1}^q n_k!}{(q - 1 + \sum_{k = 1}^q n_k)!} \right) dz_1 \dots dz_r, \\ 
        &=
    \left( \prod_{l = 1}^r \frac{b_l^{m_l+1} - a_l^{m_l+1}}{m_l + 1} \right) \left( \frac{\prod_{k = 1}^q n_k!}{(q - 1 + \sum_{k = 1}^q n_k)!} \right).
		\end{aligned}
\end{equation*}

If we adopt the convention that the settings of the process variables are rescaled to the $\left[-1, +1 \right]$ interval, the hyperrectangle becomes a hypercube and the expression for $W_{ij}$ can be simplified to
\begin{equation}
    \label{eq:wij}
    W_{ij} = \left( \prod_{l = 1}^r \frac{1^{m_l+1} - (-1)^{m_l+1}}{m_l + 1} \right) \left( \frac{\prod_{k = 1}^q n_k!}{(q - 1 + \sum_{k = 1}^q n_k)!} \right).
\end{equation}
In the event one of the $m_l$ values is odd, $1^{m_l+1} - (-1)^{m_l+1}$ is zero and $W_{ij}$ also becomes zero. In the event all $m_l$ values are even, $1^{m_l+1} - (-1)^{m_l+1}$ is equal to $2$.

So, for example, in the case where there are three mixture variables and one process variable (i.e., $q = 3$ and $r = 1$) the model expansion
$\boldsymbol{f}(\boldsymbol{a}_{js})$ is 
$(x_1, x_2, x_1x_2, x_1x_3, x_2x_3, x_1z, x_2z, x_3z, z^2)^T$. Multiplying $\boldsymbol{f}(\boldsymbol{a}_{js})$ by its transpose yields the matrix
\begin{equation*}
\boldsymbol{f}(\boldsymbol{a}_{js}) \boldsymbol{f}^T(\boldsymbol{a}_{js}) = 
\begin{bmatrix}
 x_1^2 &  x_1 x_2 &  x_1^2 x_2 &  x_1^2 x_3 &  x_1 x_2 x_3 &  x_1^2 z &  x_1 x_2 z &  x_1 x_3 z &  x_1 z^2 \\
 x_1 x_2 &  x_2^2 &  x_1 x_2^2 &  x_1 x_2 x_3 &  x_2^2 x_3 &  x_1 x_2 z &  x_2^2 z &  x_2 x_3 z &  x_2 z^2 \\
 x_1^2 x_2 &  x_1 x_2^2 &  x_1^2 x_2^2 &  x_1^2 x_2 x_3 &  x_1 x_2^2 x_3 &  x_1^2 x_2 z &  x_1 x_2^2 z &  x_1 x_2 x_3 z &  x_1 x_2 z^2 \\
 x_1^2 x_3 &  x_1 x_2 x_3 &  x_1^2 x_2 x_3 &  x_1^2 x_3^2 &  x_1 x_2 x_3^2 &  x_1^2 x_3 z &  x_1 x_2 x_3 z &  x_1 x_3^2 z &  x_1 x_3 z^2 \\
 x_1 x_2 x_3 &  x_2^2 x_3 &  x_1 x_2^2 x_3 &  x_1 x_2 x_3^2 &  x_2^2 x_3^2 &  x_1 x_2 x_3 z &  x_2^2 x_3 z &  x_2 x_3^2 z &  x_2 x_3 z^2 \\
 x_1^2 z &  x_1 x_2 z &  x_1^2 x_2 z &  x_1^2 x_3 z &  x_1 x_2 x_3 z &  x_1^2 z^2 &  x_1 x_2 z^2 &  x_1 x_3 z^2 &  x_1 z^3 \\
 x_1 x_2 z &  x_2^2 z &  x_1 x_2^2 z &  x_1 x_2 x_3 z &  x_2^2 x_3 z &  x_1 x_2 z^2 &  x_2^2 z^2 &  x_2 x_3 z^2 &  x_2 z^3 \\
 x_1 x_3 z &  x_2 x_3 z &  x_1 x_2 x_3 z &  x_1 x_3^2 z &  x_2 x_3^2 z &  x_1 x_3 z^2 &  x_2 x_3 z^2 &  x_3^2 z^2 &  x_3 z^3 \\
 x_1 z^2 &  x_2 z^2 &  x_1 x_2 z^2 &  x_1 x_3 z^2 &  x_2 x_3 z^2 &  x_1 z^3 &  x_2 z^3 &  x_3 z^3 &  z^4 
\end{bmatrix}.
\end{equation*}

To illustrate how $W_{11}$ is calculated, we start from the first element in the first row and the first column in this matrix, i.e., $x_1^2$. This term is the square of the first mixture ingredient proportion. Hence, its exponent $n_1$ is equal to $2$. The other two mixture variables, $x_2$ and $x_3$, are not present, meaning their exponents $n_2$ and $n_3$ are $0$. Additionally, this element does not involve any process variables, meaning $m_1 = 0$. Using Equation~(\ref{eq:wij}), we obtain
\begin{equation*}
    W_{11} = 
    \left(  \frac{1^{m_1+1} - (-1)^{m_1+1}}{m_1 + 1} \right) \left( \frac{n_1! \times n_2! \times n_3!}{(3 - 1 + n_1 + n_2 + n_3)!} \right) = 
    \left(  \frac{1^{0+1} - (-1)^{0+1}}{0 + 1} \right) \left( \frac{2! \times 0! \times 0!}{(3 - 1 +  2 + 0 + 0)!} \right)
    = \left( \frac{2}{1}  \right)\left(  \frac{2}{24} \right)
    = \frac{1}{6}.
\end{equation*}

As another illustration, we calculate $W_{99}$. To this end, we start from the element in the last row and the last column of $\boldsymbol{f}(\boldsymbol{a}_{js}) \boldsymbol{f}^T(\boldsymbol{a}_{js})$, i.e., $z^4$. This term is the process variable raised to the 4-th power. Hence, $m_1 = 4$. None of the mixture variables are present, meaning that their exponents are all $0$, and thus $n_1 = n_2 = n_3 = 0$.  So, using Equation~(\ref{eq:wij}) again, we obtain
\begin{equation*}
    W_{99} =  
    \left(  \frac{1^{m_1+1} - (-1)^{m_1+1}}{m_1 + 1} \right) \left( \frac{n_1! \times n_2! \times n_3!}{(3 - 1 + n_1 + n_2 + n_3)!} \right) = 
    \left(  \frac{1^{4+1} - (-1)^{4+1}}{4 + 1} \right) \left( \frac{0! \times 0! \times 0!}{(3 - 1 +  0 + 0 + 0)!} \right)
    = \left( \frac{2}{5}  \right)\left(  \frac{1}{2} \right)
    = \frac{1}{5}.
\end{equation*}

Following this process for each of the elements in the matrix $\boldsymbol{f}(\boldsymbol{a}_{js}) \boldsymbol{f}^T(\boldsymbol{a}_{js})$, we obtain the full moments matrix,
\begin{equation*}
    \boldsymbol{W} = 
    \begin{bmatrix}
        \frac{1}{6} & \frac{1}{12} & \frac{1}{30} & \frac{1}{30} & \frac{1}{60} & 0 & 0 & 0 & \frac{1}{9} \\
        \frac{1}{12} & \frac{1}{6} & \frac{1}{30} & \frac{1}{60} & \frac{1}{30} & 0 & 0 & 0 & \frac{1}{9} \\
        \frac{1}{30} & \frac{1}{30} & \frac{1}{90} & \frac{1}{180} & \frac{1}{180} & 0 & 0 & 0 & \frac{1}{36} \\
        \frac{1}{30} & \frac{1}{60} & \frac{1}{180} & \frac{1}{90} & \frac{1}{180} & 0 & 0 & 0 & \frac{1}{36} \\
        \frac{1}{60} & \frac{1}{30} & \frac{1}{180} & \frac{1}{180} & \frac{1}{90} & 0 & 0 & 0 & \frac{1}{36} \\
        0 & 0 & 0 & 0 & 0 & \frac{1}{18} & \frac{1}{36} & \frac{1}{36} & 0 \\
        0 & 0 & 0 & 0 & 0 & \frac{1}{36} & \frac{1}{18} & \frac{1}{36} & 0 \\
        0 & 0 & 0 & 0 & 0 & \frac{1}{36} & \frac{1}{36} & \frac{1}{18} & 0 \\
        \frac{1}{9} & \frac{1}{9} & \frac{1}{36} & \frac{1}{36} & \frac{1}{36} & 0 & 0 & 0 & \frac{1}{5} 
    \end{bmatrix}.
\end{equation*}

As with the D-optimality criterion, we define the Bayesian I-optimality criterion as the I-optimality criterion averaged over the prior distribution $\pi(\boldsymbol{\theta})$ of the parameter vector $\boldsymbol{\theta}$:
\begin{equation}    \label{eq_pseudo_bayes_i_eff}
    \mathcal{I}_B =
    \int_{\mathbb{R}^{m}}
    \mathrm{tr}\left[ \boldsymbol{I}^{-1}(\boldsymbol{X}, \boldsymbol{\theta}) \boldsymbol{W}\right] 
    \pi(\boldsymbol{\theta}) d\boldsymbol{\theta}.
\end{equation}

\subsection{Numerical approximation to optimality criteria}

The Bayesian optimality criteria must be approximated numerically because there is no closed-form solution to the integrals in Equations~(\ref{eq_pseudo_bayes_d_eff}) and~(\ref{eq_pseudo_bayes_i_eff}). This is usually done by using random or systematic draws from the prior distribution $\pi(\boldsymbol{\theta})$ (\cite{kessels_efficient_2009}; \cite{ruseckaite_bayesian_2017}; \cite{train_discrete_2009}; \cite{yu2010comparing}; \cite{becerra2021bayesian}). In our work, we utilize Halton draws from the prior distribution because they reduce the variance of the approximation to the integral and provide a good coverage of the entire domain of the prior distribution (\cite{train_discrete_2009}; \cite{yu2010comparing}). Moreover, \textcite{bhat2001quasi} verified that around 100 Halton draws provide about the same level of accuracy as 2000 pseudo-random draws in the context of a 5-dimensional approximation to the likelihood of a mixed multinomial model. \textcite{yu2010comparing} showed that Halton draws also produce good approximations of integrals with higher dimensions in the context of optimal design for choice experiments.
Denoting the number of Halton draws by $R$ and each individual draw by $\boldsymbol{\theta}^{(i)}$, our approximations for Equations~(\ref{eq_pseudo_bayes_d_eff}) and~(\ref{eq_pseudo_bayes_i_eff}) are
\begin{equation}    \label{eq_pseudo_bayes_d_eff_mc}
    \mathcal{D}_B \approx 
        \frac{1}{R} \sum_{i = 1}^R
        \left[  
        \det \left( 
            \boldsymbol{I}^{-1}(\boldsymbol{X}, \boldsymbol{\theta}^{(i)})
         \right) \right]^{\frac{1}{m}},
\end{equation}
and
\begin{equation}    \label{eq_pseudo_bayes_i_eff_mc}
    \mathcal{I}_B \approx
        \frac{1}{R} \sum_{i = 1}^R
        \mathrm{tr} \left[ \boldsymbol{I}^{-1}(\boldsymbol{X}, \boldsymbol{\theta}^{(i)}) \boldsymbol{W} \right],
\end{equation}
respectively. 

Like \textcite{ruseckaite_bayesian_2017} and \textcite{becerra2021bayesian}, we used $R = 128$ Halton draws from a multivariate normal prior distribution in both of our examples in the next section. We verified numerically that this number of draws provided a sufficiently good approximation of the Bayesian optimality criteria for the numbers of parameters in the models used in the two examples.

\subsection{Construction of D- and I-optimal designs}\label{algorit}

To compute our optimal designs, we used a coordinate-exchange algorithm (\cite{meyer_nachtsheim}; \cite{goos_jones_optimal_2011}). A coordinate-exchange algorithm was also used by \textcite{kessels_efficient_2009}, \textcite{ruseckaite_bayesian_2017}, and \textcite{becerra2021bayesian} in the context of choice experimentation. \textcite{becerra2021bayesian} implemented their algorithm in the R programming language \autocite{rlang} with the aid of several existing R packages (\cite{ggtern2018}; \cite{ggplot2_2016}; \cite{devtools2020}; \cite{rcpp2011}; \cite{rcpp2013}; \cite{rcpp2018}; \cite{rcppArmadillo2014}; \cite{purrr2020}), and created a package called \texttt{opdesmixr}, available at \texttt{https://github.com/mariobecerra/opdesmixr}, which allows the computation of locally D-optimal, Bayesian D-optimal, locally I-optimal, and Bayesian I-optimal designs for first-order, second-order, and special-cubic Scheffé models. We extended the package and added the functionality to compute locally D-optimal, Bayesian D-optimal, locally I-optimal and Bayesian I-optimal designs for the model presented in Equation~(\ref{eq:utility_js_pv}), involving mixture ingredient proportions as well as process variables.

The coordinate-exchange algorithm we implemented starts from a random initial design, and begins by optimizing the first ingredient proportion of the first alternative within the first choice set, followed by the second ingredient proportion of the first alternative within the first choice set, and so on, until all $q$ ingredient proportions have been optimized. Then, it continues with each of the $r$ process variables. The algorithm then repeats this process for each alternative in each choice set in the design. The whole process is repeated until the design can no longer be improved or until a maximum number of iterations has been reached. At each step of the coordinate-exchange algorithm, we seek the optimal value of every individual ingredient proportion $x_{ijs}$ or process variable setting $z_{ijs}$. This is a univariate optimization problem which can be solved in a straightforward way using Brent's univariate optimization method \autocite{brent1973algorithms}. Every time Brent's univariate optimization method is invoked during the course of the coordinate-exchange algorithm, the Bayesian D- or I-optimality criterion has to be evaluated. Despite the efficient approximation of these criteria using Halton draws, this renders the coordinate-exchange algorithm for choice experiments computationally intensive.

As indicated in \textcite{piepel_construction_2005}, \textcite{goos_jones_optimal_2011}, \textcite{ruseckaite_bayesian_2017}, and \textcite{becerra2021bayesian}, the coordinate-exchange algorithm must be modified to deal with mixtures. Since the mixture proportions must sum up to one, they cannot be independently changed. As a matter of fact, a change in one proportion requires a change in at least one other proportion. This dependency is solved by using the so-called Cox effect direction (\cite{cornell2002experiments}; \cite{goos_jones_optimal_2011}; \cite{piepel_construction_2005}). After a change of one of the ingredient proportions, $x_{ijs}$, to $x_{ijs} + \Delta$, we modify the other $q-1$ proportions as follows:
\begin{equation*}
    x_{kjs}^{\mathrm{new}} = 
    \begin{cases} 
      \left( 1 - \frac{\Delta}{1 - x_{ijs}} \right) x_{kjs}  & \mathrm{if} \quad x_{ijs} \neq 1, \\
      \frac{ 1 - (x_{ijs} + \Delta)}{q - 1} & \mathrm{if} \quad x_{ijs} = 1. 
   \end{cases}
\end{equation*}


\section{Results}\label{sec:results}

In this section, as proofs of concept, we present D- and I-optimal designs for two example choice experiments involving a mixture and one or more process variables. In both examples, we use a normal prior distribution, which is the most commonly used prior in the literature on the optimal design of choice experiments. The first example involves a cocktail tasting experiment and was inspired by \textcite{courcoux1997methode}, while the second example involves a fish patty experiment. The inspiration for this example came from \textcite{cornell2002experiments, cornell1984fractional, cornell1988analyzing} and \textcite{goos2022fish}.

\subsection{Cocktail example}\label{cex}%

\textcite{courcoux1997methode} discussed an experiment in which cocktails involving mango juice, blackcurrant syrup, and lemon juice were tasted. The experiment was conducted by asking respondents to taste different pairs of cocktails and indicating their preferred one in each pair. \textcite{ruseckaite_bayesian_2017} and \textcite{becerra2021bayesian} revisited this experiment, computed prior distributions for the parameter vector $\boldsymbol{\theta}$ of a special-cubic Scheffé model, and created optimal experimental designs using this prior.

In the experiment, \textcite{courcoux1997methode} imposed lower bounds of 0.3, 0.15 and 0.1 on the three ingredient proportions. To deal with this issue and to be able to use our implementation of the coordinate-exchange algorithm, like \textcite{ruseckaite_bayesian_2017} and \textcite{becerra2021bayesian}, we expressed the mixtures defining the cocktails in terms of so-called pseudocomponents $x_1$, $x_2$, and $x_3$. These pseudocomponents are defined such that they take a minimum value of $0$ and a maximum value of $1$, and sum up to one. The conversion of the true ingredient proportions into pseudocomponent proportions is done via the formula $x_i = (a_i - L_i)/(1 - L)$, where $L_i$ denotes the lower bound of ingredient $i$, $a_i$ denotes the true ingredient proportion, and $L$ is the sum of the lower bounds for all $q$ ingredient proportions.

In September 2019, students at KU Leuven replicated the experiment by asking $35$ respondents to taste cocktails made with mango juice, blackcurrant syrup, and lemon juice and say which one they preferred. Each respondent tasted four choice sets of two cocktails. This experiment, as the original in \textcite{courcoux1997methode}, did not have process variables. Nonetheless, since the preference for a cocktail may depend on the temperature at which it is served, we used this data and created additional simulated responses with a synthetic process variable related to temperature to obtain a prior normal distribution for parameter vector $\boldsymbol{\theta}$ using the model in Equation~(\ref{eq:utility_js_pv}).
We then fitted a multinomial logit model to these data, which gave us an estimated mean and variance-covariance-matrix, which in turn we used to construct the prior distribution in our cocktail example.
Our prior mean vector is $\boldsymbol{\theta} = (7.562, 0.907, 5.109, 14.573, 17.1806, 19.2705, 19.2705, 19.2705, 0)^T$. This means that the utility of alternative $j$ in choice set $s$ was modeled as
\begin{equation*}
    \begin{aligned}
        U_{js} &=
        7.562 x_{1js} + 0.907 x_{2js} \\
        &+ 5.109 x_{1js} x_{2js} + 14.573 x_{1js} x_{3js} + 17.1806 x_{2js} x_{3js} \\
        &+ 19.2705 x_{1js} z_{1js} + 19.2705 x_{2js} z_{1js} +  19.2705 x_{3js} z_{1js} \\
        &+ 0 z_{1js}^2 + \varepsilon_{js}.
    \end{aligned}
\end{equation*}
The prior variance-covariance matrix we used is $\boldsymbol{\Sigma}_0 = \mathrm{diag}(4, 9, 49, 36, 49, 900, 900, 900, 900)$. It must be noted that the variances in this matrix were rounded to the nearest integer from the estimated multinomial logit model. With this prior distribution, we computed Bayesian D- and I-optimal designs using the coordinate-exchange algorithm discussed in Section~\ref{algorit}.

\begin{figure}[ht]
    \centering
    \includegraphics[width=0.99\textwidth]{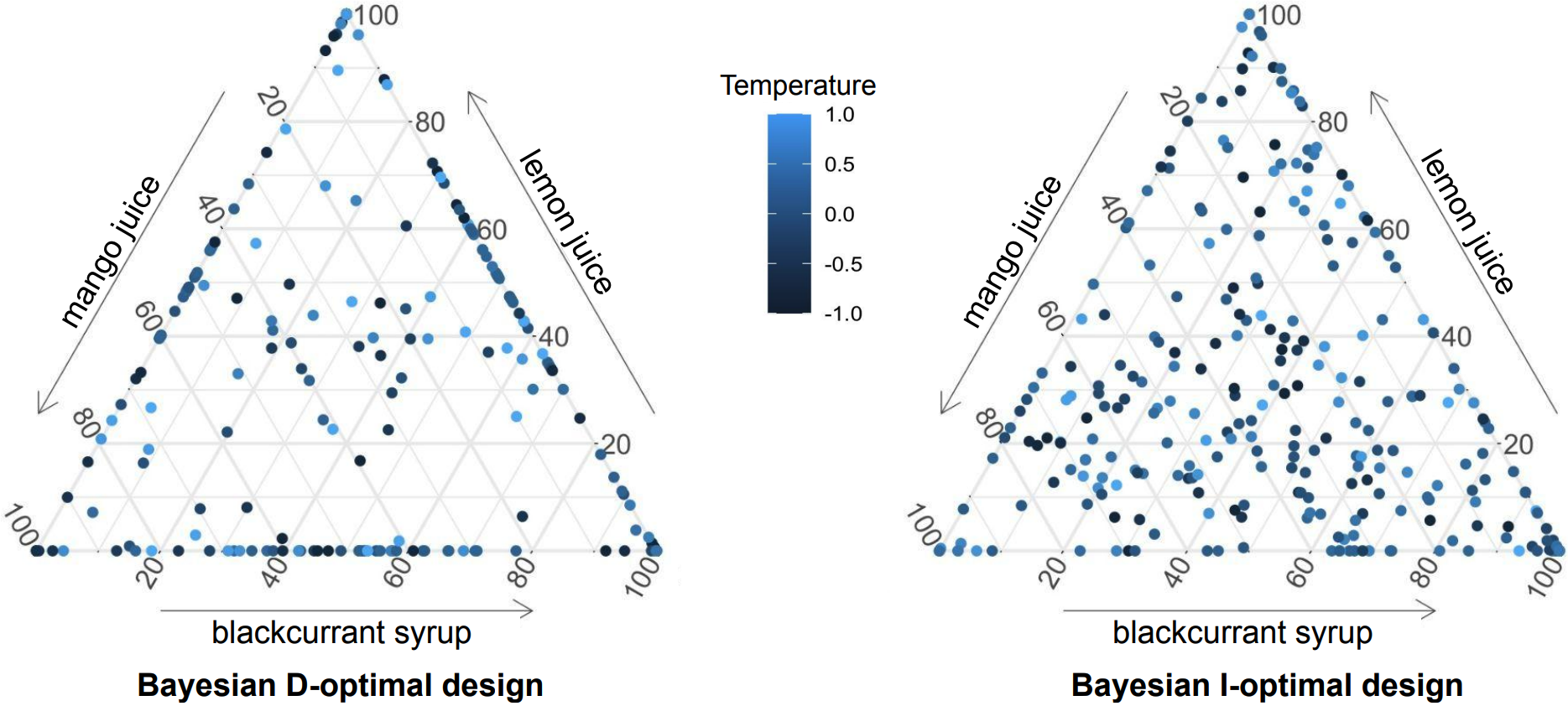}
    \caption{Bayesian D- and I-optimal designs produced by our coordinate-exchange algorithm for the cocktail experiment.}
    \label{fig:res_cocktail_db_vs_ib_design_ternary}
\end{figure}

\begin{figure}[ht]
    \centering
    \includegraphics[width=0.65\textwidth]{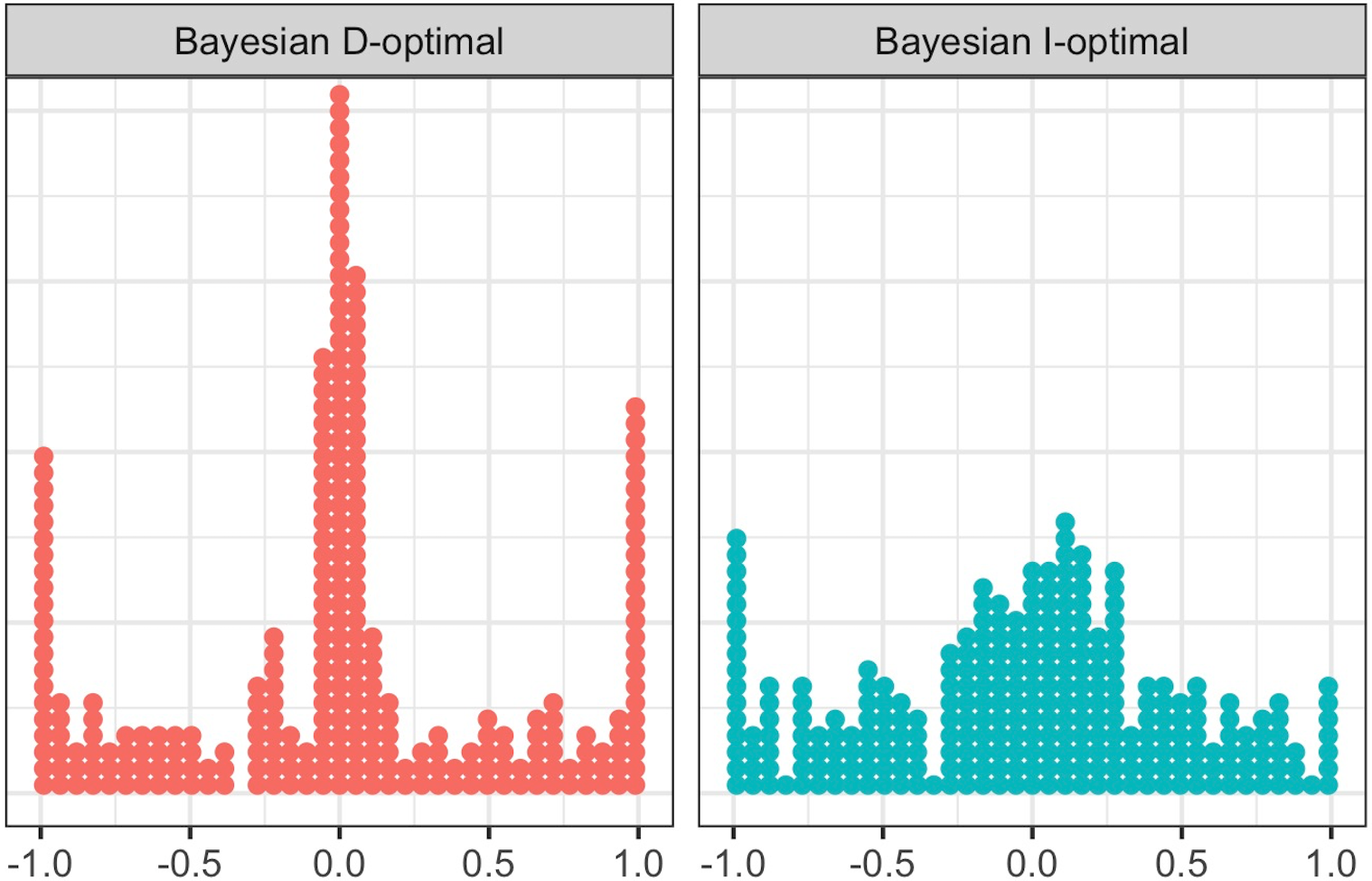}
    \caption{Distribution of the value of the process variable temperature in the Bayesian D- and I-optimal designs for the cocktail experiment.}
    \label{fig:cocktail_process_variable_dotplot}
\end{figure}

Our Bayesian D- and I-optimal designs are shown graphically in Figure~\ref{fig:res_cocktail_db_vs_ib_design_ternary}. In the figure, the mixtures in each of the $35 \times 4 = 140$ choice sets are presented in terms of the pseudocomponent proportions. The shade of blue of each dot denotes the level of process variable temperature for the corresponding mixture. Figure~\ref{fig:cocktail_process_variable_dotplot} shows the distribution of the temperatures selected for the alternatives in the 140 choice sets in each of the designs. 

In Figure~\ref{fig:res_cocktail_db_vs_ib_design_ternary}, it can be seen that the points in the I-optimal design are spread more evenly over the entire simplex compared to those of the D-optimal counterpart. This is consistent with the results of \textcite{becerra2021bayesian} for choice experiments with mixtures in the absence of process variables. It is also worth pointing out that both designs use  levels other than $-1$ and $+1$ for the process variable temperature, even though the mean prior value for the quadratic effect of the process variable temperature is zero.

\begin{figure}[ht]
    \centering
    \includegraphics[width=0.99\textwidth]{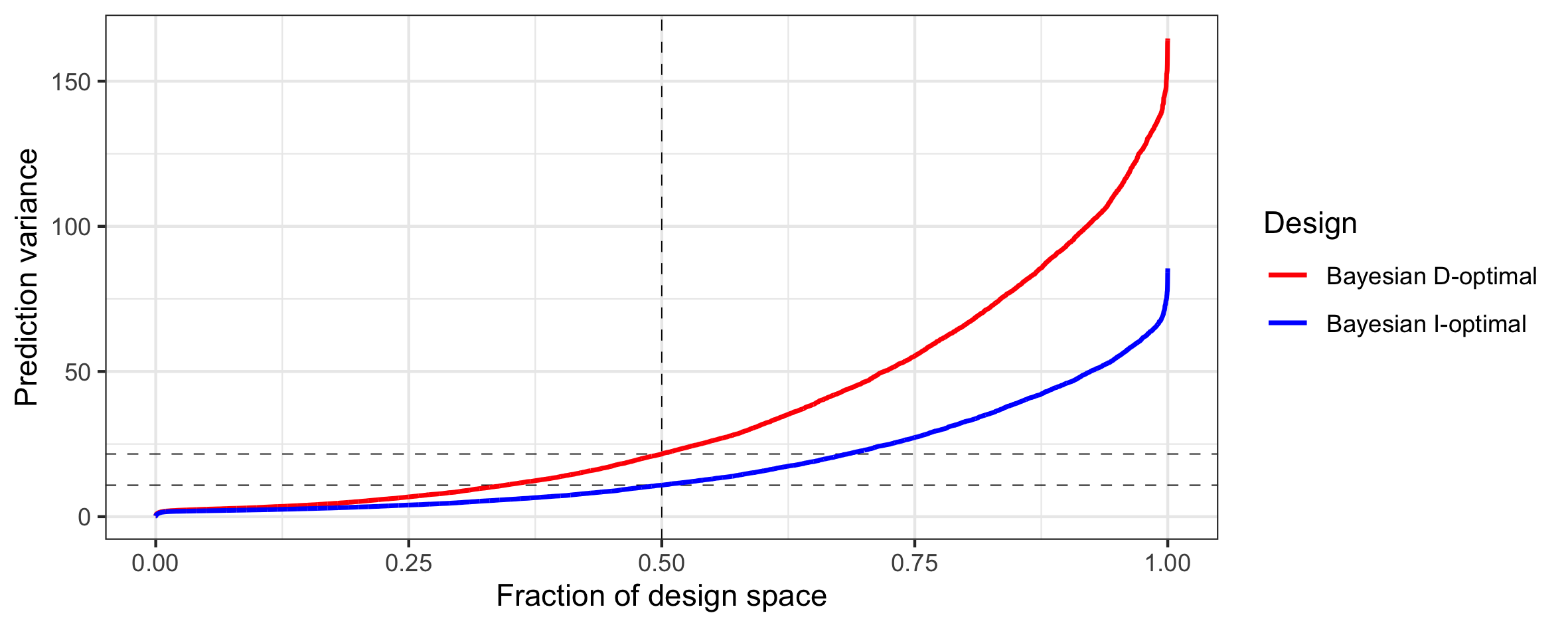}
    \caption{Fraction of design space plot of our Bayesian D- and I-optimal designs for the cocktail experiment.}
    \label{fig:res_cocktail_fds_db_vs_ib_plot}
\end{figure}

Figure \ref{fig:res_cocktail_fds_db_vs_ib_plot} shows the fraction of design space plots of the two Bayesian optimal designs. These plots display the performance of the designs in terms of the prediction variance for each point in the experimental region or design space \autocite{zahran2003fraction}. The horizontal axis corresponds to a fraction of the experimental region, while the vertical axis ranges from the minimum to the maximum prediction variance over the entire experimental region \autocite{goos_jones_optimal_2011}. A curve in a fraction of design space plot shows the prediction variances $\boldsymbol{f}^T(\boldsymbol{x}) \boldsymbol{I}^{-1}(\boldsymbol{X}, \boldsymbol{\beta}) \boldsymbol{f}(\boldsymbol{x})$ for a large number of random points selected from the experimental region, ordered from small to large. Ideally, all prediction variances are small throughout the entire experimental region, in which case the curve in the fraction of design space plot is virtually flat. Another way of explaining the fraction of design space plot is to say that it is the cumulative distribution function of the prediction variances across the experimental region, but with the positions of the two axes swapped.


The typical method to construct a fraction of design space plot for a given design is to randomly sample a large number of points $M$ (e.g., 10,000 points) inside the experimental region. Then, the prediction variance $\boldsymbol{f}^T(\boldsymbol{x}) \boldsymbol{I}^{-1}(\boldsymbol{X}, \boldsymbol{\beta}) \boldsymbol{f}(\boldsymbol{x})$ is calculated for each of these points, and all $M$ prediction variances 
are sorted from smallest to largest to obtain the empirical cumulative distribution function of the prediction variances
(\cite{ozol2005fraction}; \cite{goldfarb2004fraction}; \cite{goos_jones_optimal_2011}). If we denote the prediction variance of the $i$-th sampled point by $v_i$, then the non-decreasing curve joining the $M$ pairs $(i/M, v_i)$ forms the fraction of design space plot. A point $i/M$ on the horizontal axis of the fraction of design space plot gives the proportion of the design space that has a prediction variance less than or equal to the corresponding value $v_i$ on the vertical axis \autocite{smucker2018optimal}. In order to deal with the issue of the prediction variance depending on the unknown parameter vector, we computed prediction variances for $128$ Halton draws from the prior distribution of the parameter vector $\boldsymbol{\theta}$ and averaged the results.

The main takeaway from Figure \ref{fig:res_cocktail_fds_db_vs_ib_plot} is that the prediction variance is much higher for the Bayesian D-optimal design than for its I-optimal counterpart. The median prediction variance for the Bayesian D-optimal design is about $21.6$, while it is about $10.9$ for the Bayesian I-optimal one.

\subsection{Fish patty example}\label{fex}%

The second example we discuss involves a fish patty and was inspired by the work of \textcite{cornell2002experiments, cornell1984fractional, cornell1988analyzing}, \textcite{goos2022fish}. In the original experiment, the interest was in the firmness of patties made with a mixture of three fish species: mullet, sheepshead, and croaker. These patties were subjected to different processing conditions: oven cooking temperature (375 or 425 degrees Fahrenheit), oven cooking time (25 or 40 minutes), and deep fat frying time (25 or 40 seconds). The first three variables are mixture variables and the last three are process variables.

Since the original interest was in the firmness of the patty, no preference data is available to construct a normal prior distribution for our example. However, assuming firmness is proportional to utility, we used the original data and the model 
\begin{equation*}
    \begin{aligned}
    Y &=
        \gamma_1^0 x_{1} + \gamma_2^0 x_{2} + \gamma_3^0 x_{2}\\
        & + \gamma_{12}^0 x_{1} x_{2} + \gamma_{13}^0 x_{1} x_{3} + \gamma_{23}^0 x_{2} x_{3} \\
        & + \gamma_{1}^1 x_{1} z_{1} +\gamma_{2}^1 x_{2} z_{1} + \gamma_{3}^1 x_{3} z_{1} \\
        & + \gamma_{1}^2 x_{1} z_{2} + \gamma_{2}^2 x_{2} z_{2} + \gamma_{3}^2 x_{3} z_{2} \\
        & + \gamma_{1}^3 x_{1} z_{3} + \gamma_{2}^3 x_{2} z_{3} + \gamma_{3}^3 x_{3} z_{3} \\
        & + \alpha_{12} z_{1} z_{2} + \alpha_{13} z_{1} z_{3} + \alpha_{23} z_{2} z_{3} + \varepsilon  
    \end{aligned}
\end{equation*}
to obtain a prior point estimate for the parameter vector $\boldsymbol{\theta}$. This model is the same as the one in Equation~(\ref{eq:scheffe_second_order_pv}), but without the quadratic terms for the three process variables. The reason we did not include these quadratic effects is that, in the original experiment, the process variables were studied at two levels only. As a consequence, the quadratic effects were inestimable.
We obtained the following estimate for the parameter vector
\begin{equation*}
    \begin{aligned}
    \boldsymbol{\theta}^T &= 
    (\gamma_1^0, \gamma_2^0, \gamma_3^0, \gamma_{12}^0, \gamma_{13}^0, \gamma_{23}^0, \gamma_{1}^1, \gamma_{2}^1, \gamma_{3}^1, \gamma_{1}^2, \gamma_{2}^2, \gamma_{3}^2, \gamma_{1}^3, \gamma_{2}^3, \gamma_{3}^3, \alpha_{12}, \alpha_{13}, \alpha_{23}) \\
    &= (2.864, 1.074, 2.003, -0.974, -0.834, 0.356, 0.376, 0.106, 0.206, 0.642, 0.2, 0.403, -0.078, -0.087, -0.01, 0.027, 0.001, -0.008).
    \end{aligned}
\end{equation*}
Next, we transformed the parameter vector to the identified parameter space, as explained in Section~\ref{subsec:mixtures_and_mnl_model}. To this end, we computed 
$\gamma_1^{0*} = \gamma_1^{0} - \gamma_3^{0} = 2.864 - 2.003 = 0.861$ and $\gamma_2^{0*} = \gamma_2^{0} - \gamma_3^{0} = 1.074 - 2.003 = -0.929$. As a result, our prior model for the utility of alternative $j$ in choice set $s$ in the fish patty example is
\begin{equation*}
    \begin{aligned}
        U_{js} &=
        0.861 x_{1js} - 0.929 x_{2js} \\
        &-0.974 x_{1js} x_{2js} -0.834 x_{1js} x_{3js} + 0.356 x_{2js} x_{3js} \\
        &+ 0.376 x_{1js} z_{1js} + 0.106 x_{2js} z_{1js} + 0.206 x_{3js} z_{1js} \\
        &+ 0.642 x_{1js} z_{2js} + 0.2 x_{2js} z_{2js} + 0.403 x_{3js} z_{2js} \\
        &- 0.078 x_{1js} z_{3js} -0.087 x_{2js} z_{3js} -0.01 x_{3js} z_{3js} \\
        &+ 0.027 z_{1js} z_{2js} + 0.001 z_{1js} z_{3js} - 0.008 z_{2js} z_{3js} \\
        &+ 0 z_{1js}^2 + 0 z_{2js}^2 + 0 z_{3js}^2 + \varepsilon_{js}.
    \end{aligned}
\end{equation*}

The estimates of the parameters in the initial model were used as the means of a set of normal prior distributions with variance-covariance matrices of the form $\boldsymbol{\Sigma}_0 = \kappa \boldsymbol{I}_{21}$, where $\kappa$ is a positive scalar that controls the level of uncertainty and $\boldsymbol{I}_{21}$ is the identity matrix of size $21$. A higher value of $\kappa$ indicates a higher level of uncertainty concerning the parameter values. This structure of variance-covariance gives us a simple way to study the impact of different levels of uncertainty expressed by the prior distribution on the final design.

The variance-covariance matrix $\boldsymbol{\Sigma}_0$ corresponding to the initial $21$-parameter model must then also be transformed to the identified $20$-dimensional parameter space. This results in a new $20 \times 20$ prior variance-covariance matrix
\begin{equation*}
    \boldsymbol{\Sigma}_0^{'} = 
    \begin{pmatrix*}[r]
    2 \kappa    & \kappa  & 0       & \ldots & 0      & 0 \\
    \kappa      & 2\kappa & 0       & \ldots & 0      & 0 \\
    0           & 0       & \kappa  & \ldots & 0      & 0 \\
    \vdots      & \vdots  & \vdots  & \ddots & \vdots & \vdots \\
    0           & 0       & 0       & \ldots & \kappa & 0 \\
    0           & 0       & 0       & \ldots & 0      & \kappa
    \end{pmatrix*}.
\end{equation*}

\begin{figure}[ht]
    \captionsetup[subfigure]{aboveskip=-3pt,belowskip=-2pt}
  \begin{subfigure}[b]{0.33335\textwidth}
    \centering
    \includegraphics[width=\textwidth]{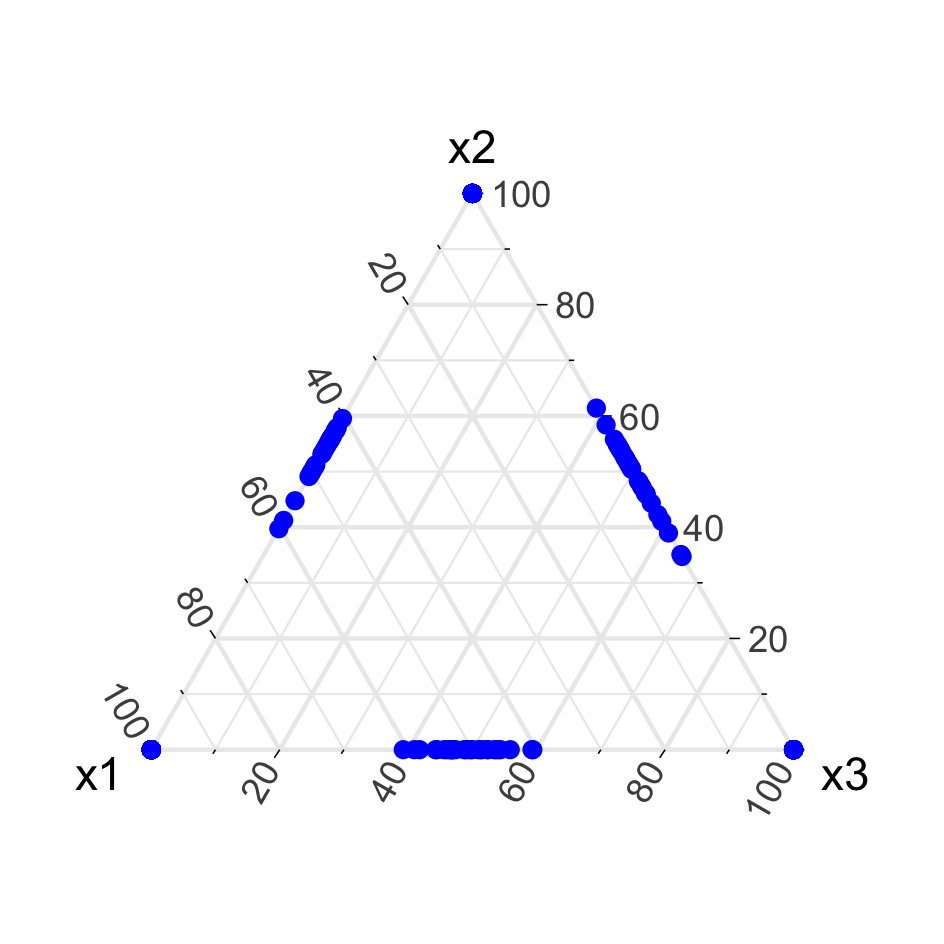}
    \caption{Bayesian D-optimal design with $\kappa = 0.5$}
    \label{fig:cornell_fishpatty_experiment_maxit10_kappa0.5_Dopt}
  \end{subfigure}
  \hfill
  \begin{subfigure}[b]{0.33335\textwidth}
    \centering
    \includegraphics[width=\textwidth, trim={13mm 1mm 25mm 15mm}, clip]{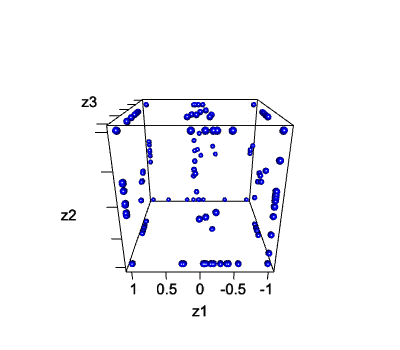}
    \caption{Bayesian D-optimal design with $\kappa = 0.5$}
    \label{fig:cornell_fishpatty_experiment_maxit10_kappa0.5_Dopt_cube_st}
  \end{subfigure}
  \hfill
\begin{subfigure}[b]{0.33335\textwidth}
    \centering
    \includegraphics[width=\textwidth]{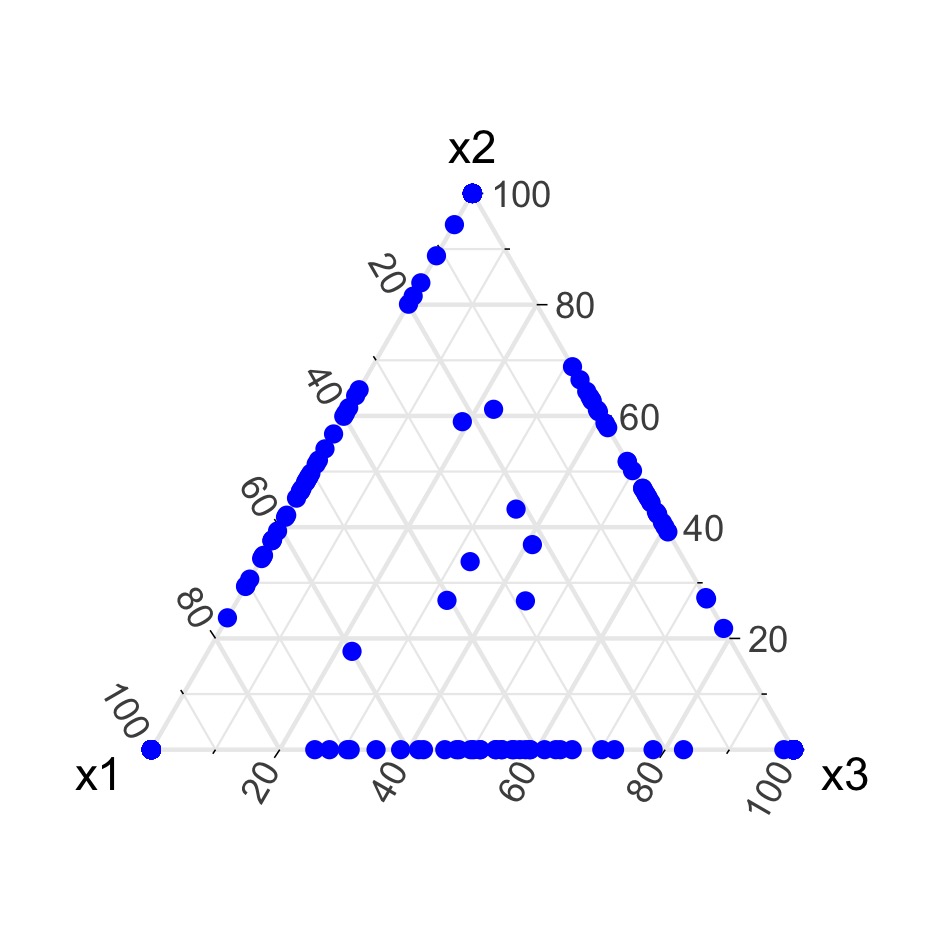}
    \caption{Bayesian D-optimal design with $\kappa = 5$}
    \label{fig:cornell_fishpatty_experiment_maxit10_kappa5_Dopt}
  \end{subfigure}
  \hfill
  \begin{subfigure}[b]{0.33335\textwidth}
    \centering
    \includegraphics[width=\textwidth, trim={13mm 1mm 25mm 15mm}, clip]{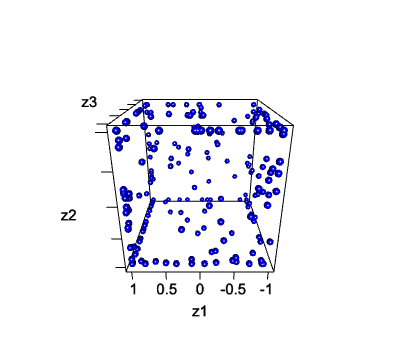}
    \caption{Bayesian D-optimal design with $\kappa = 5$}
    \label{fig:cornell_fishpatty_experiment_maxit10_kappa5_Dopt_cube_st}
  \end{subfigure}
  \hfill
  \begin{subfigure}[b]{0.33335\textwidth}
    \centering
    \includegraphics[width=\textwidth]{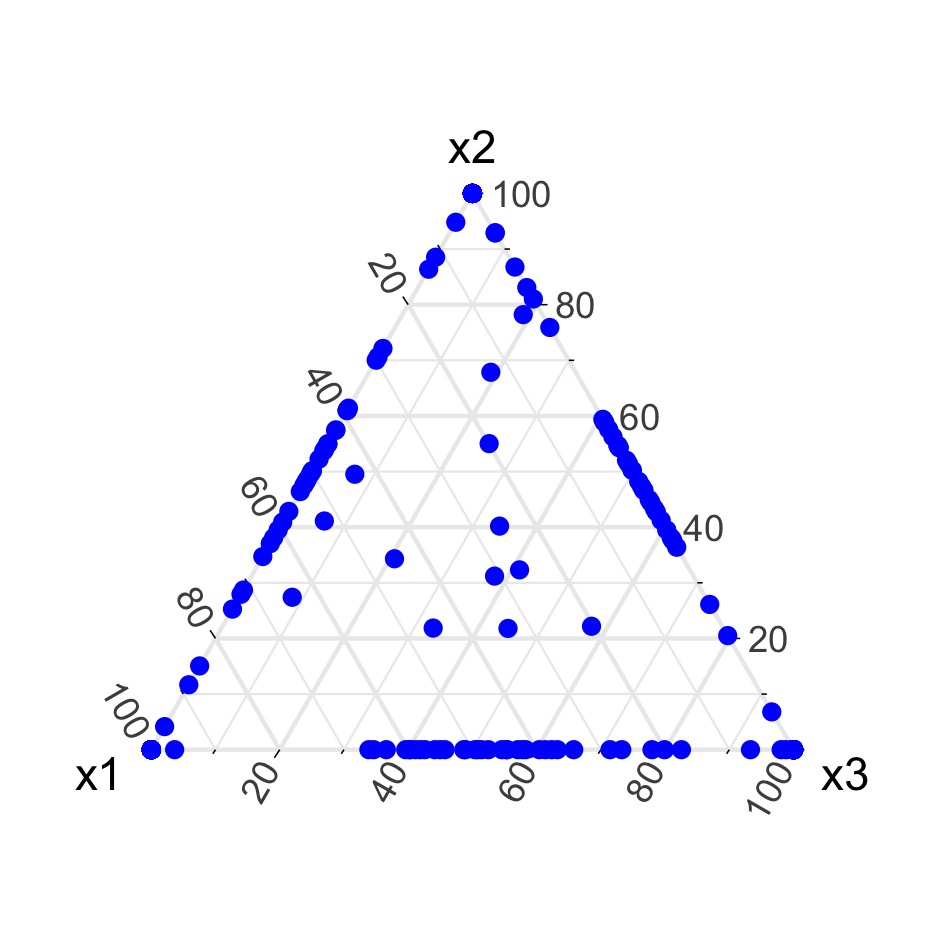}
    \caption{Bayesian D-optimal design with $\kappa = 10$}
    \label{fig:cornell_fishpatty_experiment_maxit10_kappa10_Dopt}
  \end{subfigure}
  \hfill
  \begin{subfigure}[b]{0.33335\textwidth}
    \centering
    \includegraphics[width=\textwidth, trim={13mm 1mm 25mm 15mm}, clip]{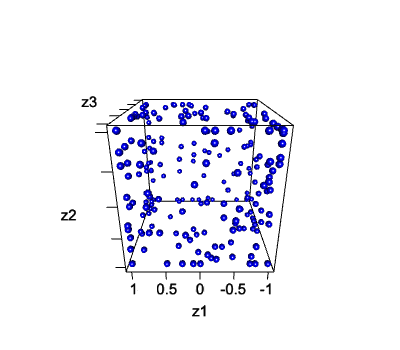}
    \caption{Bayesian D-optimal design with $\kappa = 10$}
    \label{fig:cornell_fishpatty_experiment_maxit10_kappa10_Dopt_cube_st}
  \end{subfigure}
  \hfill
  \begin{subfigure}[b]{0.33335\textwidth}
    \centering
    \includegraphics[width=\textwidth]{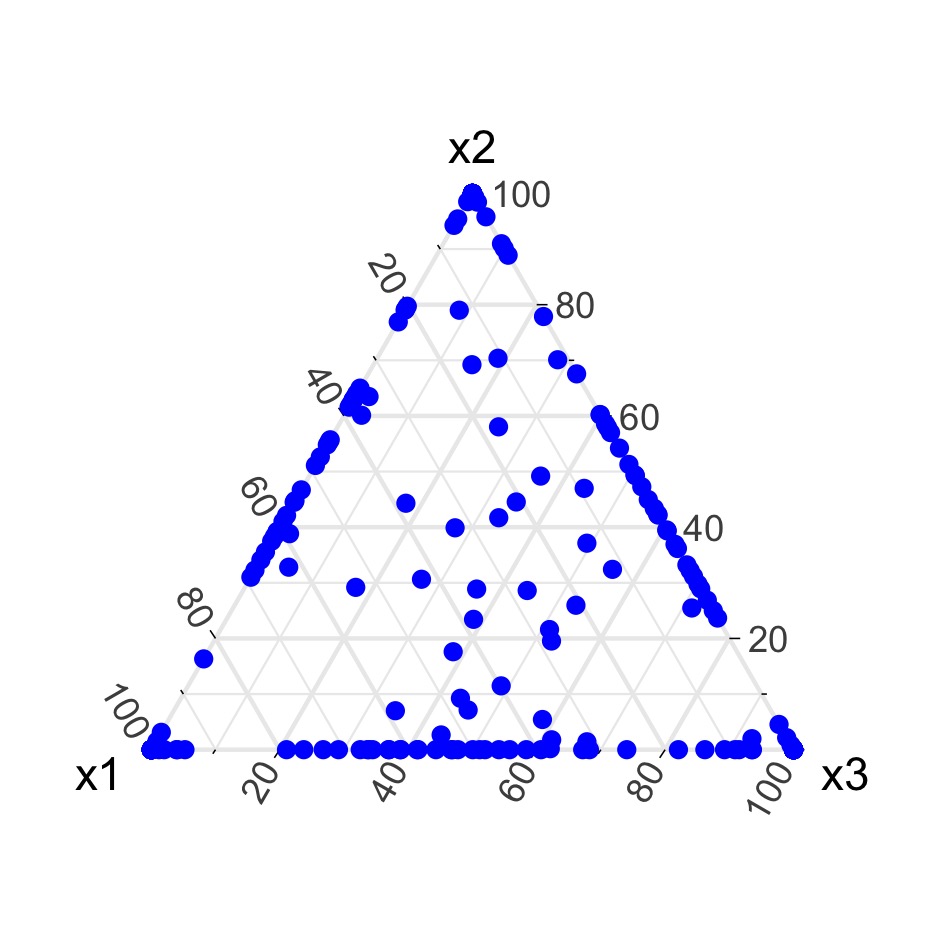}
    \caption{Bayesian D-optimal design with $\kappa = 30$}
    \label{fig:cornell_fishpatty_experiment_maxit10_kappa30_Dopt}
  \end{subfigure}
  \hfill
  \begin{subfigure}[b]{0.33335\textwidth}
    \centering
    \includegraphics[width=\textwidth, trim={13mm 1mm 25mm 15mm}, clip]{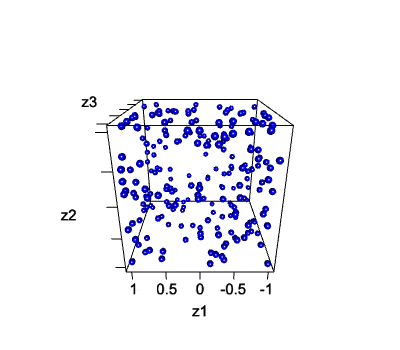}
    \caption{Bayesian D-optimal design with $\kappa = 30$}
    \label{fig:cornell_fishpatty_experiment_maxit10_kappa30_Dopt_cube_st}
  \end{subfigure}
  \hfill
  \caption{
  Bayesian D-optimal designs for the fish patty experiment. The four figures on the left show the mixture ingredient proportions, while the four figures on the right show the settings of the process variables.
  }
  \label{fig:res_cornell_ternary_Dopt}
\end{figure}

\begin{figure}[ht]
    \captionsetup[subfigure]{aboveskip=-3pt,belowskip=-2pt}
  \begin{subfigure}[b]{0.33335\textwidth}
    \centering
    \includegraphics[width=\textwidth]{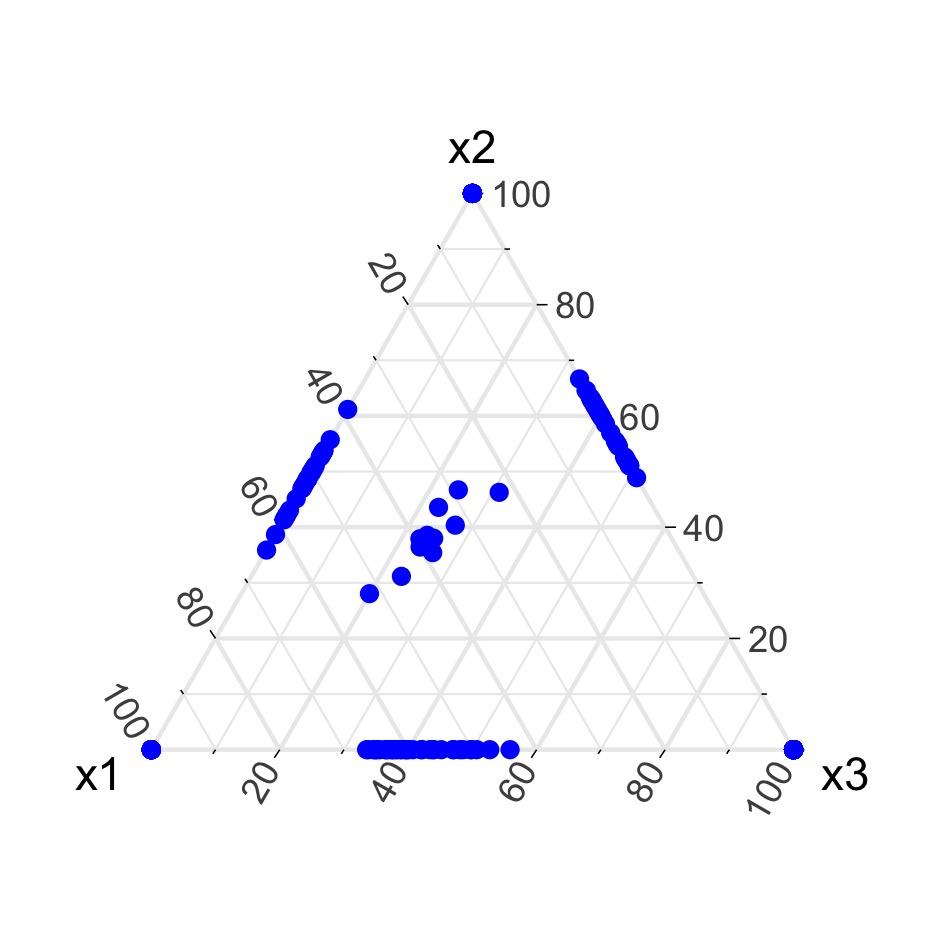}
    \caption{Bayesian I-optimal design with $\kappa = 0.5$}
    \label{fig:cornell_fishpatty_experiment_maxit10_kappa0.5_Iopt}
  \end{subfigure}
  \hfill
  \begin{subfigure}[b]{0.33335\textwidth}
    \centering
    \includegraphics[width=\textwidth, trim={13mm 1mm 25mm 15mm}, clip]{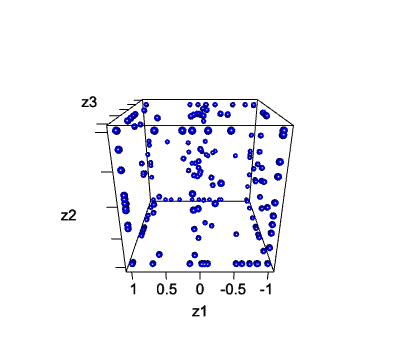}
    \caption{Bayesian I-optimal design with $\kappa = 0.5$}
    \label{fig:cornell_fishpatty_experiment_maxit10_kappa0.5_Iopt_cube_st}
  \end{subfigure}
  \hfill
\begin{subfigure}[b]{0.33335\textwidth}
    \centering
    \includegraphics[width=\textwidth]{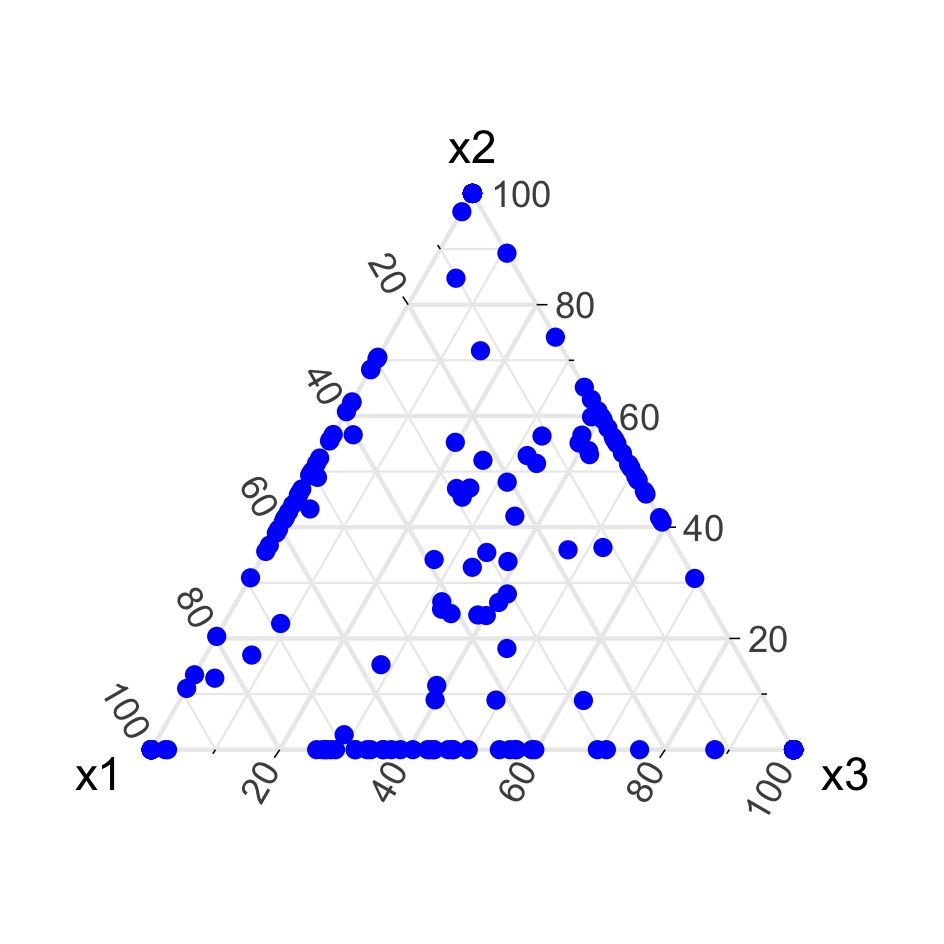}
    \caption{Bayesian I-optimal design with $\kappa = 5$}
    \label{fig:cornell_fishpatty_experiment_maxit10_kappa5_Iopt}
  \end{subfigure}
  \hfill
  \begin{subfigure}[b]{0.33335\textwidth}
    \centering
    \includegraphics[width=\textwidth, trim={13mm 1mm 25mm 15mm}, clip]{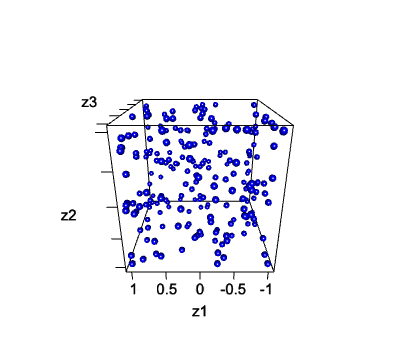}
    \caption{Bayesian I-optimal design with $\kappa = 5$}
    \label{fig:cornell_fishpatty_experiment_maxit10_kappa5_Iopt_cube_st}
  \end{subfigure}
  \hfill
  \begin{subfigure}[b]{0.33335\textwidth}
    \centering
    \includegraphics[width=\textwidth]{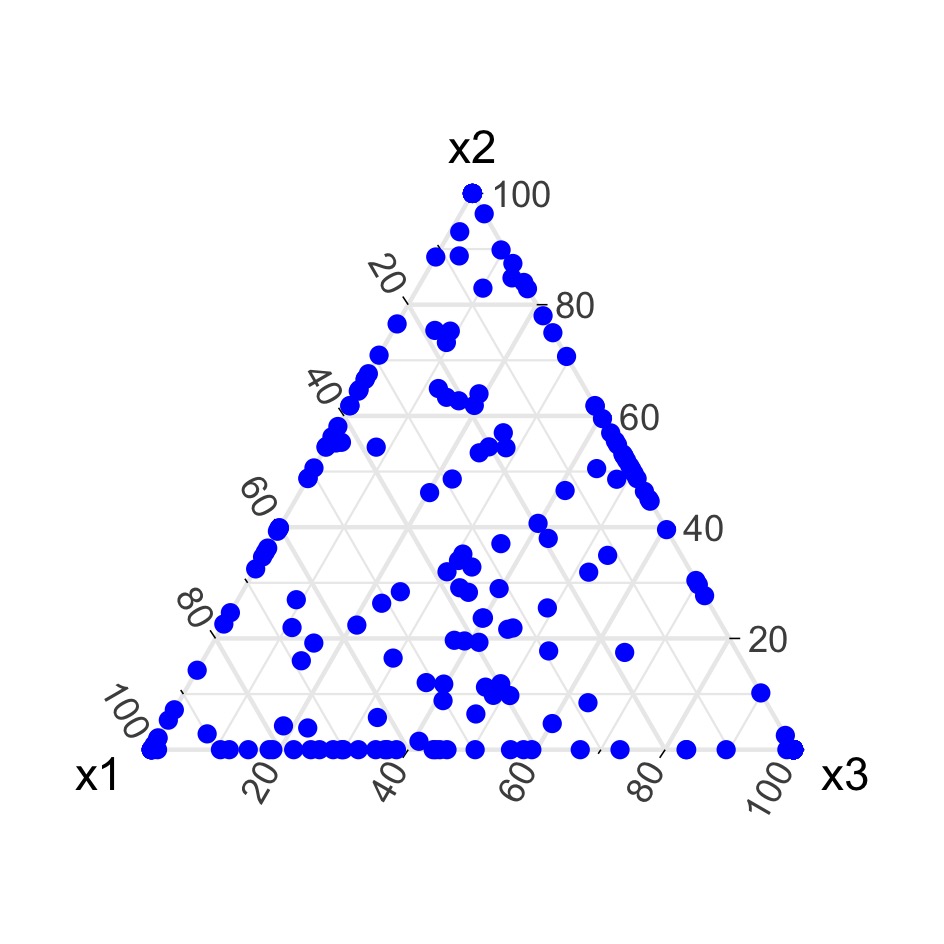}
    \caption{Bayesian I-optimal design with $\kappa = 10$}
    \label{fig:cornell_fishpatty_experiment_maxit10_kappa10_Iopt}
  \end{subfigure}
  \hfill
  \begin{subfigure}[b]{0.33335\textwidth}
    \centering
    \includegraphics[width=\textwidth, trim={13mm 1mm 25mm 15mm}, clip]{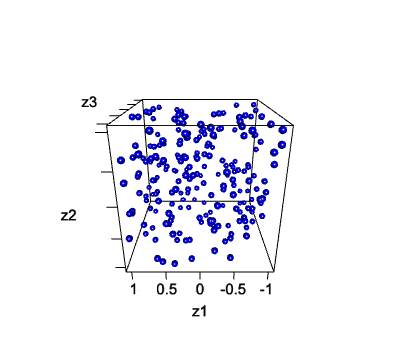}
    \caption{Bayesian I-optimal design with $\kappa = 10$}
    \label{fig:cornell_fishpatty_experiment_maxit10_kappa10_Iopt_cube_st}
  \end{subfigure}
  \hfill
  \begin{subfigure}[b]{0.33335\textwidth}
    \centering
    \includegraphics[width=\textwidth]{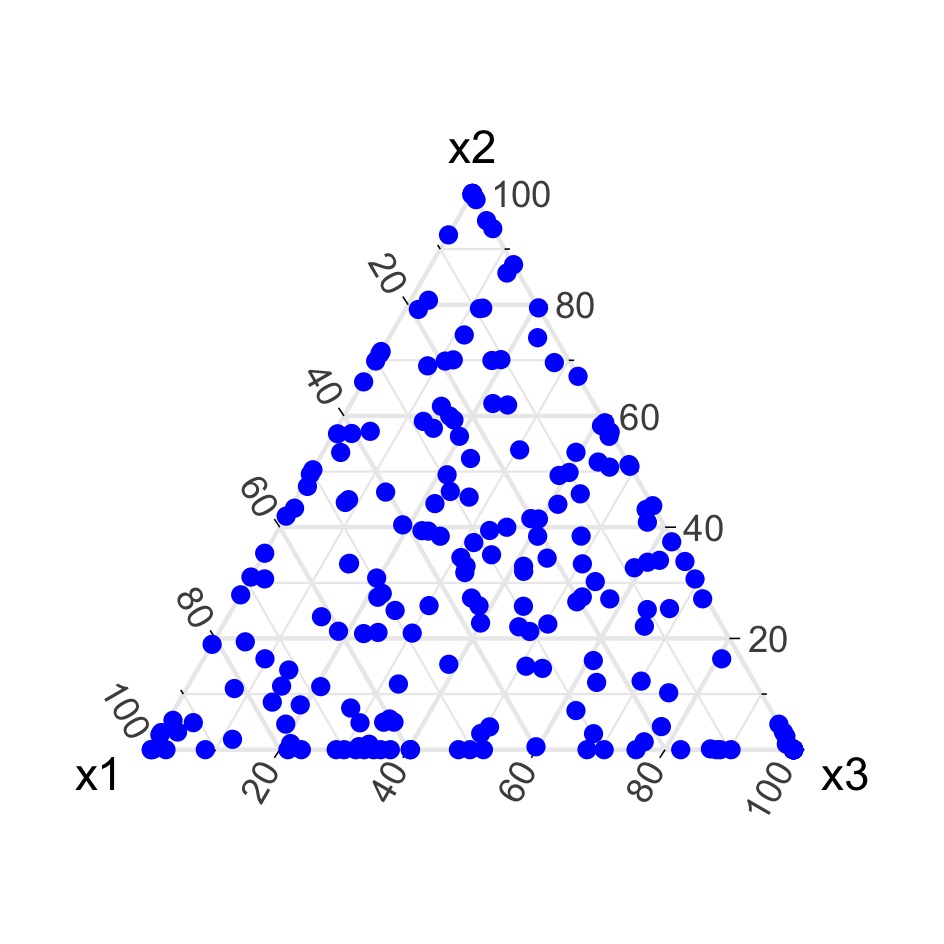}
    \caption{Bayesian I-optimal design with $\kappa = 30$}
    \label{fig:cornell_fishpatty_experiment_maxit10_kappa30_Iopt}
  \end{subfigure}
  \hfill
  \begin{subfigure}[b]{0.33335\textwidth}
    \centering
    \includegraphics[width=\textwidth, trim={13mm 1mm 25mm 15mm}, clip]{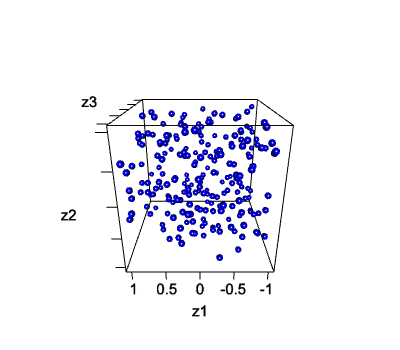}
    \caption{Bayesian I-optimal design with $\kappa = 30$}
    \label{fig:cornell_fishpatty_experiment_maxit10_kappa30_Iopt_cube_st}
  \end{subfigure}
  \hfill
  \caption{
  Bayesian I-optimal designs for the fish patty experiment. The four figures on the left show the mixture ingredient proportions, while the four figures on the right show the settings of the process variables.
  }
  \label{fig:res_cornell_ternary_Iopt}
\end{figure}

We computed Bayesian D- and I-optimal designs for the same $\kappa$ values as \textcite{ruseckaite_bayesian_2017} and \textcite{becerra2021bayesian}, that is $0.5$, $5$, $10$ and $30$. All of our Bayesian D- and I-optimal designs are shown graphically in Figures~\ref{fig:res_cornell_ternary_Dopt} and \ref{fig:res_cornell_ternary_Iopt}.
It can be seen that the spread in the points in the optimal designs increases with $\kappa$, and the spread is more pronounced for the Bayesian I-optimal designs than for the Bayesian D-optimal designs.

\begin{figure}[ht]
    \centering
    \includegraphics[width=0.99\textwidth]{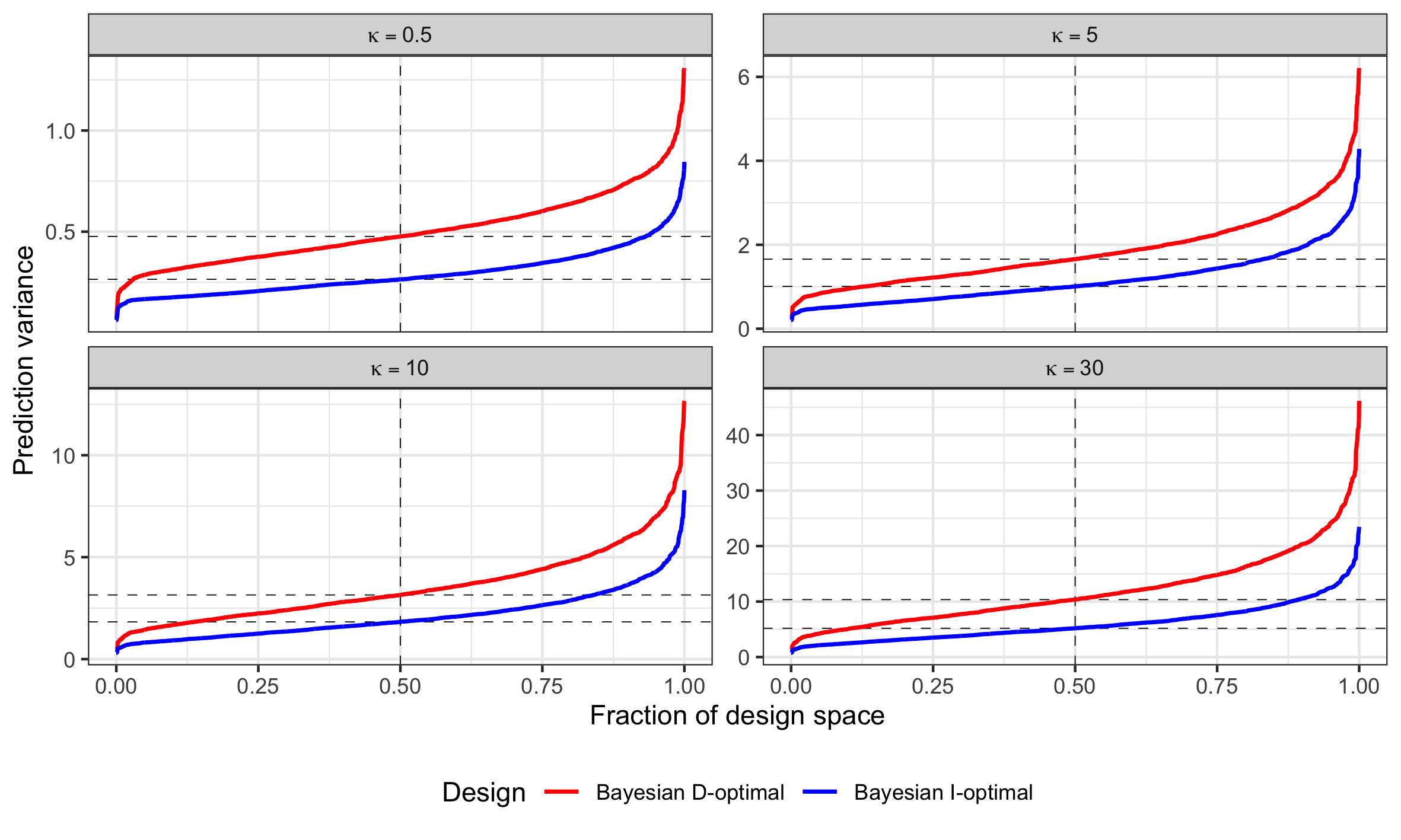}
    \caption{Fraction of design space plots of our Bayesian D- and I-optimal designs for the fish patty experiment for four values of $\kappa$, which represents the level of uncertainty concerning the prior parameter vector $\boldsymbol{\theta}$.}
    \label{fig:res_cornell_db_vs_ib_fds_plot}
\end{figure}

Figure \ref{fig:res_cornell_db_vs_ib_fds_plot} shows the fraction of design space plots for the Bayesian D- and I-optimal designs. For each value of $\kappa$, the D-optimal design has a much higher prediction variance than its I-optimal counterpart. Hence, the Bayesian I-optimal designs add substantial value in terms of precision of prediction when compared to Bayesian D-optimal designs.

\section{Discussion}\label{sec:discussion}

We introduced the theory for choice experiments involving mixtures and process variables, and embedded the Bayesian D- and I-optimality criteria in a coordinate-exchange algorithm for constructing optimal designs for this type of choice experiments. We also showed two examples in which the I-optimal designs perform substantially better than their D-optimal counterparts in terms of the variance of the predicted utility, which is something desirable because it is crucial to have precise predictions for any combination of ingredient proportions and process variables when optimizing the formulation of a mixture and the settings of the related process variables.

We identified three possible extensions of our work. The first possibility is inspired by a practical difficulty that arises when conducting choice experiments with mixtures with or without process variables. When the number of distinct mixtures appearing in the Bayesian optimal designs is large and the mixtures have to be tasted, it is logistically very complicated to perform the experiment. For instance, for a given number of tasters, organizing a choice experiment in which 40 distinct mixtures have to be tasted in perhaps 80 different choice sets is much harder to organize and perform than a choice experiment in which only 20 distinct mixtures have to be tasted in 40 different choice sets. While the former experiment may be preferable from a statistical viewpoint, it may be practically infeasible. Therefore, it is valuable to develop an algorithm that finds optimal designs with mixtures and process variables with an upper bound on the number of distinct mixtures and/or an upper bound on the number of distinct choice sets, as well as an upper bound on the number of distinct settings and values that the process variables can take.

Second, we focused on the multinomial logit model, which assumes that there is homogeneity in the preferences of the respondents. This works well in many practical scenarios, but it might be an unrealistic assumption, as demonstrated by \textcite{courcoux1997methode} and \textcite{goos_hamidouche_2019_choice}. Hence, it would make sense to extend the algorithms presented here to other types of choice models that take into account the possible presence of consumer heterogeneity, such as the mixed logit model and the latent class choice model.

A third topic for future research would be to modify our coordinate-exchange algorithm, so that it can also cope with experimental regions for the ingredient proportions that are not a simplex. Such experimental regions arise when there are constraints on the ingredient proportions other than lower bounds for individual proportions. Methodologically speaking, this is not highly innovative, since the mixture coordinate-exchange algorithm of \textcite{piepel_construction_2005} for linear regression models is able to deal with this complication. However, embedding this capability in our implementation of the coordinate-exchange algorithm for choice experiments with mixtures would be useful for practitioners.

%

Finally, we would like to point out that the work we presented here has applications in other fields of research than food. This is because choice experiments involving mixtures are relevant in, for example, transportation and economics too. As a matter of fact, \textcite{zijlstra2019mixture} conducted a choice experiments in which the mixtures between which the respondents had to choose were different ways in which a given mobility budget could be spent. \textcite{khademi2013traveler} discuss a choice experiment involving a mixture of road toll, congestion pricing and parking price. \textcite{boonaert2021twofold} use a choice experiment concerning the desired composition of a family, where the family composition is considered a mixture of boys and girls with different education levels. Finally, \textcite{yang2016prevalence} use a mixture choice experiment to measure context-dependent responses to accumulative energy charges under budget constraints. In all of these non-food-related choice experiments, an ad-hoc experimental design was used and there was a variable related to the total amount of the mixture. This total amount can be viewed as a process variable, and, therefore, the models and the optimal design approach we present here would be applicable to these choice experiments too.





\printbibliography


\end{document}